\def\bea{\begin{eqnarray}}
\def\eea{\end{eqnarray}}

\def\beq{\begin{equation}}
\def\eeq{\end{equation}}

\def\squareforqed{\hbox{\rlap{$\sqcap$}$\sqcup$}}
\def\qed{\ifmmode\squareforqed\else{\unskip\nobreak\hfil
\penalty50\hskip1em\null\nobreak\hfil\squareforqed
\parfillskip=0pt\finalhyphendemerits=0\endgraf}\fi}

\def\<{\langle}
\def\>{\rangle}

\documentstyle[prb,twocolumn,aps,rotate,epsf,floats,amssymb,amsmath]{revtex}

\newcommand{\bq}{{\boldsymbol q}}
\newcommand{\bk}{{\boldsymbol k}}
\newcommand{\bl}{{\boldsymbol l}}

\newcommand{\bP}{{\boldsymbol P}}

\newcommand{\br}{{\boldsymbol r}}

\newcommand{\bz}{{\boldsymbol z}}

\newcommand{\bX}{{\boldsymbol X}}
\newcommand{\bh}{{\boldsymbol h}}
\newcommand{\bd}{{\boldsymbol d}}

\newcommand{\bOmega}{{\boldsymbol \Omega}}
\newcommand{\obOmega}{{\overline{\boldsymbol \Omega}}}

\newcommand{\ophi}{{\overline{\phi}}}

\newcommand{\ox}{{\overline{x}}}
\newcommand{\oy}{{\overline{y}}}
\newcommand{\ovr}{{\overline{r}}}

\newcommand{\obr}{{\overline{\boldsymbol r}}}

\newcommand{\vF}{{v_{\rm F}}}
\newcommand{\ovF}{{\overline{v}_{\rm F}}}
\newcommand{\ov}{{\overline{v}}}

\newcommand{\onu}{{\overline{\nu}}}
\newcommand{\oP}{{\overline{\rm P}}}
\newcommand{\tbP}{{\tilde{\boldsymbol{P}}}}

\def\rbx#1{\raisebox{-0.3ex}{$\scriptstyle #1$}}

\begin{document}
\setlength{\unitlength}{1cm}
\renewcommand{\arraystretch}{1.4}

\title{Low temperature electronic properties of Sr$_2$RuO$_4$ II: \\
       Superconductivity}  

\author{Ralph Werner}
\address{Physics Department, Brookhaven National
Laboratory, Upton, NY 11973-5000, USA and \\ Institut f\"ur Theorie der 
Kondensierten Materie, Universit\"at Karlsruhe, 76128 Karlsruhe, Germany} 

\date{\today}

\maketitle

\centerline{Preprint. Typeset using REV\TeX.}

\begin{abstract} 
The body centered tetragonal structure of Sr$_2$RuO$_4$ gives rise to
umklapp scattering enhanced inter-plane pair correlations in the
$d_{\rm yz}$ and $d_{\rm  zx}$ orbitals. Based on symmetry arguments,
Hund's rule coupling, and a bosonized description of the in-plane
electron correlations the superconducting order parameter is found to
be a orbital-singlet spin-triplet with two spatial components. The
spatial anisotropy is 7\%. The different components of the order
parameter give rise to two-dimensional gapless fluctuations. The phase
transition is of third order. The temperature dependence of the pair
density, specific heat, NQR, Knight shift, and susceptibility are in
agreement with experimental results.   
\end{abstract}
\pacs{PACS numbers: 63.20.Kr, 75.10 Jm, 75.25 +z}


\section{Introduction}

Sr$_2$RuO$_4$ has quickly triggered a large interest because of its
structural similarity to the cuprates and the unconventional
superconductivity below $T_{\rm c}\sim 1$ K.\cite{MHY+94} The
material is tetragonal at all temperatures.\cite{BRN+98} The three
bands cutting the Fermi level with quasi two-dimensional Fermi
surfaces\cite{Oguc95,MJD+96} can be mainly associated with the three
$t_{2g}$ orbitals of the Ru$^{4+}$ ions.\cite{SCB+96,MRS01} The
transport is Fermi liquid like for $T_{\rm c} < T < 30$
K\cite{MYH+97,MIM+98,IMK+00} and strongly anisotropic along the $c$
axis.\cite{MHY+94} The enhanced specific heat, magnetic susceptibility
and electronic mass indicate the presence of significant
correlations.\cite{MHY+94,MJD+96,PSKT98}  

The superconducting order parameter was proposed to have $p$-wave
symmetry analogous to superfluid $^3$He.\cite{RS95,Bask96} This was
supported by magnetic experimental
probes,\cite{IKA+97,IMK+98,DHM+00,LFK+98} but has not been proven
unambiguously in spite of large efforts in
Andreev spectroscopy,\cite{LGL+00,MNJ+01} thermal
conductivity,\cite{TNM+01,TSN+01,ITY+01,STM+02} and 
ac-susceptibility\cite{YAM+01} measurements. No indication for
ferromagnetic correlations has been
found.\cite{SBB+99,DLS+00,SDL+01,MPS00} The specific
heat,\cite{NMF+98,NMM99,NMM00} thermal conductivity,\cite{STK+02} and
nuclear quadrupole resonance (NQR)\cite{IMK+00} are consistent with
two-dimensional gapless fluctuations in the superconducting phase. For
a more detailed overview see Refs.\
\onlinecite{MRS01,Wern02a,Wern02d,Wern02e}.  

The present paper is part II of a series of three. In part~I (Ref.\
\onlinecite{Wern02a}) the quasi one-dimensionality of the kinetic energy
of the $d_{zx}$ and $d_{yz}$ electrons has been used to derive a
bosonized effective model. At intermediate coupling the interaction
leads to a quasi one-dimensional model for the magnetic degrees of
freedom. The resulting spectrum of elementary excitations allows for
the coherent description of the observed enhanced dynamical magnetic
susceptibility\cite{SBB+99} including the observed scale
invariance.\cite{BSB+02} Together with the two-dimensional
correlations the normal state effective mass
renormalization,\cite{BJM+00,BMJ+02} the specific heat
coefficient\cite{NMF+98,NMM99,NMM00} and the static magnetic
susceptibility enhancements\cite{MHY+94,MYH+97,PSKT98} can be
described consistently.

This work is devoted to the superconducting phase. The phase
transition to the superconducting state in Sr$_2$RuO$_4$ is three
dimensional and the specific heat has a shape that is similar to that
of a mean-field transition at the transition
temperature.\cite{NMF+98,NMM00} The low temperature specific heat data
are clearly inconsistent with the BCS theory.\cite{ZR01,ALGW01} This
scenario suggests the combination of a mean-field induced
three-dimensional transition (Sec.\ \ref{sectionmf}) with electronic
correlations dominated by the quasi two-dimensional configuration
(Sec.\ \ref{sectionmicro}). Similar approaches have been successfully
applied to coupled spin chains.\cite{ETD97} The model consistently can
account for the experimental data concerning the temperature
dependence of the pair density, specific heat, susceptibility and
thermal conductivity (Sec.\ \ref{sectionapplic}). A detailed
comparison of the present approach with the $p$-wave picture is given
(Sec.\ \ref{sectionDisc}). 

Part III (Ref.\ \onlinecite{Wern02c}) consistently
explains the experimentally observed unconventional transitions under
magnetic fields in terms of the model derived here.


\section{Mean-field approach}\label{sectionmf} 


The generic model Hamiltonian is 
\begin{equation}\label{genericH}
H=\sum_{{\bl},{\bl'}\atop\nu,\nu',\sigma} t^{\nu,\nu'}_{{\bl},{\bl}'} 
         c^{\dagger}_{{\bl},\nu,\sigma} 
         c^{\phantom{\dagger}}_{{\bl}',\nu',\sigma}
+ \sum_{{\bl},\nu,\sigma\atop\nu',\sigma'} U^{\nu,\nu'}_{\sigma,\sigma'}\,
        n_{{\bl},\nu,\sigma} n_{{\bl},\nu',\sigma'}.
\end{equation}
In this notation the electron creation and annihilation operators 
are $c^{\dagger}_{{\bl},\nu,\sigma}$ and
$c^{\phantom{\dagger}}_{{\bl},\nu,\sigma}$ for orbital $\nu$ with
spin $\sigma$ on site $\bl$, $n_{{\bl},\nu,\sigma}$ is the usual
electronic density operator, $t^{\nu,\nu'}_{{\bl},{\bl}'}$ is the
hopping matrix element. The intra-orbital Coulomb
repulsion is larger than the inter-orbital repulsion, i.e., $U^{\nu =
\nu'}_{\sigma\neq\sigma'} = U_0 > U_1 = U^{\nu \neq \nu'}_{\sigma \neq
\sigma'}$ and $U_0 > U_2 = U^{\nu \neq \nu'}_{\sigma =
\sigma'}$. The Hamiltonian in Eq.\ (\ref{genericH}) is quasi
two-dimensional.\cite{Wern02a}  

The correlations in effective one-dimensional systems show power
law behavior.\cite{SCP98} They are always more singular than
two-dimensional correlations which diverge at most 
logarithmically.\cite{Schu95} Since the kinetic energy of the $d_{zx}$
and $d_{yz}$ electrons is quasi one-dimensional we expect their
correlations to play a dominant role.\cite{Wern02a} 

In order for the system to undergo a finite temperature phase
transition the coupling of the RuO$_2$ planes is necessary.\cite{MW66}
The inter-plane $d_{z\nu}$-$d_{z\nu}$ hopping has been estimated in
Ref.\ \onlinecite{Wern02a} to be $t_\perp\approx 0.02$ eV and is an
order of magnitude smaller than the in-plane hopping. The inter-plane
hopping involving $d_{xy}$ orbitals is expected to be significantly
smaller for geometry reasons.\cite{MRS01} It follows that the
$d_{z\nu}$-$d_{z\nu}$ subsystem should drive the transition. This
conclusion is in contrast to most other
approaches\cite{MRS01,ZR01,ARS97,EMJB02,KS02,OO02b} which assume the
pair correlations of the $d_{xy}$ electrons to be dominant.

The sensitivity of the superconducting instability on the
dimensionality can be probed by experiments under pressure. Recent
ultrasonic measurements\cite{OSM+02} suggest that the superconducting
$T_{\rm c}$ increases upon uniaxial pressure along the
crystallographic $c$ axis. This is reminiscent of an increase in the
inter-plane coupling and is consistent with a pairing mechanism
induced by inter-plane pair correlations as proposed by
Baskaran.\cite{Bask96}    

On the other hand $T_{\rm c}$ has been shown to decrease
under hydrostatic pressure.\cite{SMN+97,BMJ+02} This can be understood
in terms of the decrease of the in-plane lattice parameter which in
turn increases the in-plane hopping parameters which lowers the
relative interaction strength and renders the $d_{z\nu}$-$d_{z\nu}$ 
subsystem more two-dimensional. It can thus be concluded that 
hydrostatic pressure lowers the in-plane correlations\cite{Wern02a}
including those of Coopers pairs leading to a lower $T_{\rm c}$. Note
that this assumption finds strong support in the decrease of the
effective thermodynamic electronic masses under hydrostatic pressure
observed in dHvA measurements.\cite{BMJ+02}

\subsection{Normal phase instability}\label{sectionnormal}

In summary the discussion above suggests the presence of an inter-plane 
pair-correlation term between the planar $d_{z\nu}$-$d_{z\nu}$
subsystems driving the observed phase transition. Such a term can be
written in form of the lowest order coherent inter-plane pair-hopping
term:  
\begin{equation}\label{Htperp2}
H_p = \sum_{\nu,\mu,\sigma \atop \nu',\mu',\sigma'}
          V^{\mu,\mu'}_p
  \sum_{\langle \bl, \bl'\rangle}\
         c^{\dagger}_{{\bl},\mu,\sigma} 
         c^{\phantom{\dagger}}_{{\bl}',\nu,\sigma}\,
         c^{\dagger}_{{\bl},\mu',\sigma'} 
         c^{\phantom{\dagger}}_{{\bl}',\nu',\sigma'}\,.
\end{equation}
The orbital indices $\nu,\mu,\nu',\mu' \in \{x,y\}$ are restricted by the
Pauli principle for $\sigma' = \sigma$ to $\mu' \neq \mu$ and $\nu'
\neq \nu$. The second sum runs over nearest-neighbor sites where $\bl$
and $\bl'$ are on neighboring planes.

Note that such an inter-plane pair-correlation term may have different
origins. Possible suggestions are Coulomb interactions screened on the
scale of the inter-plane distance $c/2$ or electron-phonon interaction
where the phonon degrees of freedom\cite{Wern99} have been integrated
out. Since a discussion of the specific origin of the term would not
be conclusive in the context of the present approach it is
omitted. 

Instead, a phenomenological estimate for the pair-hopping amplitude
can be obtained by requiring its value to be consistent with the
single particle inter-plane hopping $t_{\perp}$, i.e., $V^{\mu,\mu'}_p
\sim t_{\perp}^2 / v_{\rm F}$. The energy denominator is estimated by
a typical in-plane energy scale. Here the kinetic scale $v_{\rm F}$
was chosen for the sake of definiteness. Depending on the specific
mechanism, an interaction energy scale $U^{\mu,\mu'}_{\sigma,\sigma'}$
as the energy denominator is also conceivable. Since the system has
been shown to be in the intermediate coupling\cite{Wern02a} regime,
i.e., $v_{\rm F} \sim U^{\mu,\mu'}_{\sigma,\sigma'}$, the numerical
outcome is similar. Considering that $t_\perp\approx 0.02$
eV\cite{Wern02a} and $v_{\rm F}=\ovF/a \approx 0.7$ eV\cite{MPS00} one
has $t_{\perp}^2 / v_{\rm F} \approx 6\ {\rm K} \ll v_{\rm F}$
suggesting that a mean-field decoupling of Eq.\ \ref{Htperp2} is
reasonable.  

Fourier transforming the Fermi operators in Eq.\ (\ref{Htperp2}) leads
to the Hamiltonian 
\begin{equation}\label{HtperpFT}
H_p =\!\!\!\! \sum_{\nu,\mu,\sigma \atop \nu',\mu',\sigma'}
  \sum_{\bk, \bk', \bq}\! \frac{V_{\bq}}{N}
         c^{\dagger}_{{\bk},\mu,\sigma} 
         c^{\dagger}_{{\bq-\bk},\mu',\sigma'} 
         c^{\phantom{\dagger}}_{{\bq-\bk'},\nu',\sigma'}\,
         c^{\phantom{\dagger}}_{{\bk'},\nu,\sigma}
\end{equation}
with the effective Potentials
\begin{equation}\label{Vq}
V_{\bq} \approx \frac{t_{\perp}^2}{v_{\rm F}}\ 
     \cos \frac{a}{2}q_x \cos \frac{a}{2}q_y \cos \frac{c}{2}q_z\,.
\end{equation}
For $\bq = \bq_j$ with  
\begin{equation}\label{q_i}
\bq_j \in \left\{
\left(\frac{2\pi}{a}, \frac{2\pi}{a}, \frac{2\pi}{c} \right)^\dagger, 
\frac{2\pi}{a}\ \hat{x}, \frac{2\pi}{a}\ \hat{y}, \frac{2\pi}{c}\ \hat{z}
           \right\}
\end{equation}
the potential is extremal and attractive, i.e., $V_{\bq_j} \approx
-t_\perp^2/v_{\rm F} < 0$. As a consequence of the body 
centered tetragonal lattice structure the $\bq_j$ are reciprocal
lattice vectors. In other words, the body centered tetragonal lattice
symmetry allows for an um\-klapp scattering driven pair
instability.

Within the configuration space of the $d_{zx}$-$d_{yz}$ 
subsystem six possible pairing states can be found. There is an 
``equal-flavor'' doublet which can be identified as a spin-singlet 
\begin{equation}\label{efpair}
P_{0,\pm}(\bq)=\sum_{\bk}\left(
         c^{\phantom{\dagger}}_{{\bq-\bk},x,\uparrow}\,
         c^{\phantom{\dagger}}_{{\bk},x,\downarrow}   \pm
         c^{\phantom{\dagger}}_{{\bq-\bk},y,\uparrow}\,
         c^{\phantom{\dagger}}_{{\bk},y,\downarrow}\right) ,
\end{equation}
as well as a ``mixed-flavor'' doublet with a spin-singlet and a
$S^z=0$ spin-triplet component 
\begin{equation}\label{mfs0pair}
P_{1,\pm}(\bq) = \sum_{\bk}\left(
         c^{\phantom{\dagger}}_{{\bq-\bk},x,\uparrow}\,
         c^{\phantom{\dagger}}_{{\bk},y,\downarrow}   \pm
         c^{\phantom{\dagger}}_{{\bq-\bk},x,\downarrow}\,
         c^{\phantom{\dagger}}_{{\bk},y,\uparrow}\right) . 
\end{equation}
The ``mixed-flavor'' doublet with spin-triplet components with
magnetization quantum number $|S^z|=1$ is
\begin{equation}\label{mfs1pair}
P_{2,\pm}(\bq)=\sum_{\bk}\left(
         c^{\phantom{\dagger}}_{{\bq-\bk},x,\uparrow}\,
         c^{\phantom{\dagger}}_{{\bk},y,\uparrow}     \pm
         c^{\phantom{\dagger}}_{{\bq-\bk},x,\downarrow}\,
         c^{\phantom{\dagger}}_{{\bk},y,\downarrow}\right).
\end{equation}

The six components can be regrouped\cite{Wern02e} as an orbital-triplet
spin-singlet  
\begin{equation}
{\bP^{t}_{s}}^\dagger = 
       \left(P^\dagger_{0,+},P^\dagger_{0,-},P^\dagger_{1,-}\right)
\end{equation}
and a orbital-singlet spin-triplet
\begin{equation}
{\bP_{t}^{s}}^\dagger = 
       \left(P_{2,+}^\dagger,P^\dagger_{2,-},P^\dagger_{1,+}\right)
\end{equation}
given here in vectorial representation.\cite{Notationquote} The pair
operator components have a very similar structure as the $p$-wave
components initially discussed by Rice and Sigrist.\cite{RS95}
Notably, the $P_{2,\pm}$ components have the same spin-structure as
the unitary state with $E_u$ symmetry proposed in the $p$-wave
scenario for Sr$_2$RuO$_4$.\cite{SAF+99} The degrees of freedom form a 
two-dimensional analog to those of the $A$ phase\cite{VK83,Volo92} of
$^3$He. The crucial difference is that all components $P_{a,\pm}$
render even parity real-space wavefunctions because the antisymmetry
of the total electronic wavefunction is established through either the 
spin-singlet or the orbital-singlet configuration. A closer account of
the resulting symmetries is given separately in Ref.\
\onlinecite{Wern02e}. 

Introducing the variables $a\in\{0, 1, 2\}$ and $b\in\{+,-\}$ and
switching to time-dependent pair operators in the Heisenberg
representation the bare pair susceptibilities are  
\begin{equation}\label{chiPabnull}
\chi^{(P)}_{\rbx{a,b}}(\bq,\omega)=
     \langle P^{\phantom\dagger}_{a,b}(\bq,\omega)
                   P^\dagger_{a,b}(\bq,\omega) \rangle\,.
\end{equation}
The random phase approximation (RPA) is equivalent in its static limit
to the mean-field approximation. The RPA pair susceptibility is 
\begin{equation}\label{chiPsfRPA}
\chi_{\rbx{P}}(\bq,\omega)=\frac{1}{4}\sum_{a,b}
     \frac{\chi^{(P)}_{\rbx{a,b}}(\bq,\omega)}
          {1 + V_{\bq}\ 
                       \chi^{(P)}_{\rbx{a,b}}(\bq,\omega)}\,.
\end{equation}
The mean-field phase transition occurs at $T=T_{\rm c}$ when the
Stoner criterion $\chi^{(P)}_{\rbx{a,b}}(\bq_j,0)|_{T_{\rm c}} =
\chi^{(P)}_{\rbx{a,b}}(0,0)|_{T_{\rm c}} = -V_{\bq_j}^{-1} =
V_{0}^{-1} $ is fulfilled. The estimated value of $V_{0} \approx 6$ K
is consistent with the low critical temperature of $T_{\rm c}=1.5$ K.  

In order to make a quantitative prediction of $T_{\rm c}$ the
$\chi^{(P)}_{\rbx{a,b}}(0,0)$ have to be determined as a function of
temperature within the effectively two-dimensional model based on Eq.\
(\ref{genericH}). This is not straightforward as becomes clear from
the discussions in Ref.\ \onlinecite{Wern02a} and is left to future
studies. 

Mean-field approaches do not allow to control rigorously the local
constraint imposed by the Pauli exclusion principle.\cite{LW97} Since
the filling factors of the orbitals\cite{MPS00} are $n_{\mu}\sim 1.3$
it is important in a mean-field approach to include effects of the
Pauli exclusion principle at least in an approximate manner. Formally
this can be achieved  
by introducing a phenomenological orbital-dependent weighing factor
$g^2_{a}$ of order one in the denominator of Eq.\ (\ref{chiPsfRPA}):   
\begin{equation}\label{RPAwithg}
\left[1 + V_{\bq}\ \chi^{(P)}_{\rbx{a,b}}(\bq,\omega) 
                            \right]^{-1}  \to 
\left[1 + g^2_{a}\ V_{\bq}\ \chi^{(P)}_{\rbx{a,b}}(\bq,\omega)
                            \right]^{-1} .
\end{equation}
The pair-hopping process in Eq.\ (\ref{Htperp2}) with $\mu=\mu'$,
i.e., $a=0$ in Eq.\ (\ref{RPAwithg}), requires the orbital $\mu$ to be
unoccupied in the initial state which is very unlikely with
$n_{\mu}\sim 1.3$. Pair-hopping processes with $\mu\neq\mu'$ in Eq.\
(\ref{Htperp2}), where $a=1$ or $a=2$ in Eq.\ (\ref{RPAwithg}), are
much less effected by the Pauli exclusion principle. Consequently
$g^2_{0} \ll g^2_{1} \approx g^2_{2} \sim 1$ and the phase transition
occurs in the mixed flavor channels.\cite{Taki00} This conclusion is
only altered if the in-plane equal-flavor-doublet correlations
$\chi^{(P)}_{\rbx{0,b}}(\bq,\omega)$ are much larger than those of the
mixed-flavor sector for which there is no indication to be found. 

In the presence of Hund's rule coupling the spin-triplet energy
expectation values  are always smaller than those of the
spin-singlet.\cite{Bask96} 
\begin{equation}
|\langle {\bP^{s}_{t}}^\dagger  |\, H \,| 
                             {\bP^{s}_{t}}^\dagger \rangle| 
 <  
|\langle {\bP^{t}_{s}}^\dagger  |\, H \,| 
                             {\bP^{t}_{s}}^\dagger \rangle| 
\end{equation}
Consequently $\langle P_{0,+}^{\dagger} \rangle = \langle
P_{0,-}^{\dagger} \rangle = \langle P_{1,-}^{\dagger} \rangle =
0$. Consistent with $S = 1$ moments on Ru$^{4+}$ impurities in
Sr$_2$IrO$_4$,\cite{CBK+94} one finds mixed-orbital spin-triplet
pairing or, equivalently, orbital-singlet spin-triplet pairing. The
initial SU(2)$\otimes$SU(2) symmetry of the two electron spins is
broken down to SO(3). The order parameter carries spin one and
explains the absence of a change of the Knight shift\cite{IMK+98} and
of the magnetic susceptibility\cite{DHM+00} in the superconducting
phase.  

At this point it is important to notice that the estimated value for
Hund's rule coupling\cite{LL00} in Sr$_2$RuO$_4$ of $J_{\rm H} \approx
0.2 - 0.4$ eV is larger than the estimate for the spin-orbit
coupling\cite{NS00} of $\lambda \approx 0.1$ eV. Consequently
orbital-singlet pairing is possible even if the degeneracy of the
$d_{zx}$ and  $d_{yz}$ orbitals is lifted due to spin-orbit coupling
because the larger Hund's rule coupling over-compensates the effect.


\subsection{Comment on competing energies}\label{sectionenergies}

Since in the context of the superconducting cuprates it has become
habitual to discuss the pair instability in terms of the competition
between kinetic and potential energies\cite{Hirs02} it is useful here
to clarify the situation in the present approach to Sr$_2$RuO$_4$. 

In the superconducting state of a BCS superconductor the single
particle excitations are gapped and hence the system looses kinetic
energy in all spatial directions. The loss is over-compensated by the
gain in condensation energy. The intuitive idea is that the system
gains ``potential energy'' trough the attractive potential mediated by
the phonons. 

Here the situation is similar, the system looses kinetic
energy both in-plane as well as out-of-plane. In the light of the
two-dimensionality of the band structure\cite{MRS01} the loss of
kinetic energy in the RuO$_2$ planes is at least an order of magnitude
larger then that along the $c$ axis. The gain in pair hopping 
energy that one might anticipate through the inter-plane coupling term
Eq.\ (\ref{Htperp2}) is quite small---unlike the case
discussed\cite{CEKO02} in the superconducting cuprates.   

Instead, the gain in condensation energy over-compensates the loss of
kinetic energy as becomes obvious from the sine-Gordon model onto
which part of the degrees of freedom in Sr$_2$RuO$_4$ are mapped in
Sec.\ \ref{sectionsG}: the ground state energy is lowered proportional
to a given power of the gap {\em even though} the single particle
excitations are gapped [c.f.\ Eq.\ \ref{Fmu0} and Ref.\
\onlinecite{LZ97}]. Or, viewed from the perspective of the collective
condensation, the ground state energy is lowered {\em because} the 
single particle excitations are gapped. 

Moreover, the notion of potential and kinetic energy is obscured in
the bosonized approach introduced in Ref.\ \onlinecite{Wern02a} and
applied here in Sec.\ \ref{sectionmicro}. Notably in the
one-dimensional limit with preserved SU(2) invariance the Bose fields
and their dual fields are equivalent\cite{Emer79,Sene99,CT02} and so are
kinetic and potential energy. In this light the discrimination between
the two appears obsolete and the energy gain is explained most
naturally simply by the condensation of the electrons in the
collective state below the gap.


\subsection{Superconducting phase}\label{sectionsuper}

The order parameter is a spin-one SO(3) triplet with components
$\langle P^s_{t,\nu} \rangle \neq 0$. Together with the global phase
degree of freedom this implies that the to order parameter has
SO(3)$\otimes$SO(2) symmetry.

These results are consistent with the implications from the bosonized
microscopic model of the in-plane correlations discussed in Sec.\
\ref{sectionmicro}. There it will be shown that there are two
degenerate saddle points with slightly different spatial
anisotropies. This leads to a total SO(3)$\otimes$SO(2)$\otimes$SO(2)
symmetry of the order parameter. The thermodynamic implications of the
resulting gapless modes of the order parameter are discussed in
detail in Sec.\ \ref{sectionfluct}.

The usual mean-field decoupling of the pair operators is 
\begin{equation}\label{mfdecoupling}
{\bP_{t}^{s}}^\dagger{\bP_{t}^{s}}^{\phantom{\dagger}} \approx
 {\bP_{t}^{s}}^\dagger \langle {\bP_{t}^{s}}^{\phantom{\dagger}} \rangle
+    {\bP_{t}^{s}}^{\phantom{\dagger}}
                        \langle {\bP_{t}^{s}}^{\phantom{\dagger}} \rangle^*
-    |\langle {\bP_{t}^{s}}^{\phantom{\dagger}} \rangle|^2\,.
\end{equation}
The mean-field decoupling Eq.\ (\ref{mfdecoupling}) has to be applied
to the inter-plane Hamiltonian Eq.\ (\ref{Htperp2}). The mean-field
contribution $H_{\rm mf}\approx H_{\perp}$ is 
\begin{equation}\label{Hmf}
H_{\rm mf} = V_{0} \sum_{\bl} \
      \left[2\,{\rm Re}\, {\bP_{t}^{s}}^\dagger(\bl) 
              \langle {\bP_{t}^{s}}^{\phantom{\dagger}} \rangle -
\left| \langle {\bP_{t}^{s}}^{\phantom{\dagger}} \rangle \right|^2\right] .
\end{equation}
The total Hamiltonian $H + H_{\rm mf}$ usually permits a Landau
expansion of the free energy of the system. This is done in Sec.\
\ref{sectionFree} consistent with the bosonized microscopic model for
the in-plane correlations to allow for a more quantitative analysis.

An important critique of the orbital-singlet pairing mechanism is that 
in the {\em absence of interaction} electrons can only pair where the
idealized, one-dimensional bands of the $d_{zx}$ and $d_{yz}$ orbitals
cross in reciprocal space. The resulting pairing phase space is much
too small to account for the specific heat anomaly at the phase
transition [c.f. Sec.\ \ref{sectionC}]. At {\em intermediate coupling}, 
as relevant for Sr$_2$RuO$_4$,\cite{Wern02a} interaction effects
significantly enhance the pairing phase space. A close discussion is
given in Ref.\ \onlinecite{Wern02e} where it is shown that the
interactions lead to a rather homogeneous gap function. See also Sec.\
\ref{sectionFree}.


\subsection{$\gamma$ sheet: inter-band proximity effect}\label{sectiongamma}

The coupling of the different bands through the inter-band proximity
effect and the resulting single transition temperature for all
electronic degrees of freedom has been discussed in earlier
work.\cite{MRS01,ZR01,EMJB02} In the present model the driving 
correlations are those of the $d_{zx}$ and $d_{yz}$ electrons as a
result of the geometrically strongly enhanced inter-plane coupling
of with respect to the $d_{xy}$ electrons.\cite{MRS01,Wern02a} 

To obtain a qualitative theoretical picture including the $d_{xy}$
orbitals one must consider the terms in the full Hamiltonian Eq.\
(\ref{genericH}) that are left out of the $d_{zx}$-$d_{yz}$
Hamiltonian Eqs. (\ref{Hc(m)}) and (\ref{Hintz}) discussed in Sec.\
\ref{sectionmicro}. 
\begin{eqnarray}\label{Hgamma}
H_\gamma &=& \sum_{{\bl},{\bl'},\sigma} 
   t^{\gamma,\gamma}_{{\bl},{\bl}'} 
         c^{\dagger}_{{\bl},\gamma,\sigma} 
         c^{\phantom{\dagger}}_{{\bl}',\gamma,\sigma}
+ \sum_{{\bl},\sigma} U^\gamma_0\,
        n_{{\bl},\gamma,\uparrow} n_{{\bl},\gamma,\downarrow}.
\nonumber\\&&\hspace{0ex}
+\ \sum_{{\bl},\sigma \atop \nu\neq\gamma} \left( U^\gamma_1\,
        n_{{\bl},\gamma,\sigma} n_{{\bl},\nu,\sigma'\neq\sigma}.
+  U^\gamma_2\, n_{{\bl},\gamma,\sigma} n_{{\bl},\nu,\sigma}\right)
\end{eqnarray}
The hopping parameters where given in Ref.\ \onlinecite{Wern02a} as  
$t^{\gamma,\gamma}_{{\bl},{\bl}} = -0.39$ eV, 
$t^{\gamma,\gamma}_{{\bl},{\bl}+\hat{x}} =
t^{\gamma,\gamma}_{{\bl},{\bl}+\hat{y}} = -0.27$ eV, and 
$t^{\gamma,\gamma}_{{\bl},{\bl}+\hat{x}+\hat{y}} = -0.11$ eV. From the
symmetry of the $t_{2g}$ orbitals\cite{MRS01} follows that in lowest
order the interaction parameters $U^\gamma_0$, $U^\gamma_1$, and
$U^\gamma_2$ are similar to those introduced in Ref.\
\onlinecite{Wern02a} in the context of the $d_{zx}$-$d_{yz}$ subsystem
and are thus smaller but of the similar magnitude as $v_{\rm F}$. 

The inter-orbital terms $\sim U^\gamma_1$ and $\sim U^\gamma_2$ can
formally be mean-field decoupled via
\begin{eqnarray}\label{gammamf}
n_{{\bl},\gamma,\sigma} n_{{\bl},\nu,\sigma'} &\approx&
   \langle c^{\dagger}_{{\bl},\gamma,\sigma} 
           c^{\dagger}_{{\bl},\nu,\sigma'} \rangle\
   c^{\phantom{\dagger}}_{{\bl},\gamma,\sigma}
   c^{\phantom{\dagger}}_{{\bl},\nu,\sigma'}
\nonumber\\&&\hspace{5ex}
+\
   \langle c^{\phantom{\dagger}}_{{\bl},\gamma,\sigma}
           c^{\phantom{\dagger}}_{{\bl},\nu,\sigma'} \rangle\
   c^{\dagger}_{{\bl},\gamma,\sigma} 
   c^{\dagger}_{{\bl},\nu,\sigma'}
\nonumber\\&&\hspace{5ex}
-\   
   \langle c^{\dagger}_{{\bl},\gamma,\sigma}
           c^{\dagger}_{{\bl},\nu,\sigma'} \rangle
   \langle c^{\phantom{\dagger}}_{{\bl},\gamma,\sigma}
           c^{\phantom{\dagger}}_{{\bl},\nu,\sigma'} \rangle\,.
\end{eqnarray}
The mean-field contributions $\langle c_{{\bl},\gamma,\sigma}
c_{{\bl},\nu,\sigma'}\rangle$ couple the $d_{xy}$ electrons
directly to the pair instability driven by the correlations in the
$d_{zx}$ and $d_{yz}$ bands and induce pairs in the $\gamma$
band. There is only one phase transition. 

Since the inter-band proximity effect is usually strong\cite{ZR01} a
single gap is assumed herein. It has been argued that Sr$_2$RuO$_4$
might represent a particular case where the inter-band proximity
effect is suppressed\cite{ARS97} leading to a double gap\cite{KS02}
structure. No signature of a second gap has been observed in
point contact\cite{LGL+00} or thermal conductivity
experiments\cite{STK+02} though.

In principle the inter-plane pair hopping term Eq.\ (\ref{Htperp2})
must be extended to include the $d_{xy}$ or $\gamma$ orbitals, i.e.,
$\nu,\mu,\nu',\mu' \in \{x,y,\gamma\}$ as well as the free energy Eq.\
(\ref{FLandau}). Since from the orbital geometry $g_{\gamma,\gamma}
\ll g_{\gamma,x} = g_{\gamma,y} \ll g_{x,y}$  those contributions are
small so they will only contribute small quantitative corrections.


\section{Microscopic in-plane model}\label{sectionmicro}  

The microscopic model derived for the $d_{zx}$-$d_{yz}$ subsystem in
Ref.\ \onlinecite{Wern02a} successfully has been applied to describe
the normal state properties in Sr$_2$RuO$_4$. Consequently it is
reasonable to adapt the same approach for the in-plane correlations to
describe the superconducting properties. To assure consistency in the
notation of the present manuscript the elementary results are given in
the following. Please refer to Ref.\ \onlinecite{Wern02a} for the
details. 

As discussed in Sec.\ \ref{sectionmf} the relevant correlations are
those of the electrons in the $d_{zx}$ and $d_{yz}$ orbitals. The dominant
hopping amplitudes are quasi one-dimensional and given by $t_0 =
t^{x,x}_{{\bl},{\bl} + \hat{x}} = t^{y,y}_{{\bl},{\bl} +
\hat{y}}$ in Eq.\ (\ref{genericH}). The bands are linearized with
Fermi velocity $\vF \approx \sqrt{3}\ t_0$. To study the qualitative
properties of the model with parameters relevant for Sr$_2$RuO$_4$ it
proves useful to express the two spin and the two orbital degrees of
freedom trough the charge ($\phi_{\rho}$), spin ($\phi_{\rm s}$),
flavor ($\phi_{\rm f}$), and spin-flavor ($\phi_{\rm sf}$) Bose fields
and their conjugate momenta ($\Pi_\mu$). The representation can be
simplified by rotating the reference frame through
$\ox=\frac{1}{\sqrt{2}}(x+y)$ and $\oy=\frac{1}{\sqrt{2}}(x-y)$ with
$\obr=(\ox,\oy)^\dagger$. The charge (magnetic) sector Hamiltonian
including forward scattering terms is given by
\begin{eqnarray}\label{Hc(m)}
H_{\rm c(m)} &=& \lim_{L\to \infty}\frac{1}{2} \int_{-L}^L\!\! d^2\ovr\,
    \Big\{\vF \left(\Pi^2_{\rho({\rm s})} + \Pi^2_{\rm f(sf)}\right) 
\nonumber\\&&\hspace{5ex}
 +\ V_{\rm c(m)}\! \left[\partial_{\ox} \phi_{\rho({\rm s})}
    + \partial_{\oy} \phi_{\rm f(sf)}\right]^2
\nonumber\\&&\hspace{5ex}
 +\ \overline{V}_{\rm\!\!c(m)}\! 
                   \left[ \partial_{\oy} \phi_{\rho({\rm s})}
    + \partial_{\ox} \phi_{\rm f(sf)}\right]^2
\Big\}\,,
\end{eqnarray} 
with 
\begin{eqnarray} \label{Vc}
V_{\rm c(m)} &=& \vF+(-)U_0+[U_1+(-)U_2]\,,
\\               \label{oVc}
\overline{V}_{\rm c(m)} &=& \vF+(-)U_0-[U_1+(-)U_2]\,.
\end{eqnarray}
Equation (\ref{Hc(m)}) is the Hamiltonian of a crossed sliding
Luttinger liquid.\cite{MKL01,KKGA02} 

In the limit $\overline{V}_{\rm m} \ll V_{\rm m}$ the magnetic
correlations exhibit dominant one-dimensional contributions along the
diagonals of the RuO$_2$ planes.\cite{Wern02a} On a mean-field level
the spin ($\mu = {\rm s}$, $\onu = \oy$) and spin-flavor ($\mu = {\rm
  sf}$, $\onu = \ox$) channels decouple.   
\begin{equation}\label{effHmu}
H_{\mu}= \frac{v_{\mu} L}{2}\int_{} d\onu
    \left[K_{\mu}^{}\Pi^2_{\mu} + K_{\mu}^{-1}          
                         (\partial_\onu \phi_{\mu})^2\right]\,. 
\end{equation}
It is then possible to approximate $H_{\rm m} \approx H_{\rm s} +
H_{\rm sf}$. The Luttinger liquid parameter $K_{\mu}$ and the velocity
$v_{\mu}$ are effective parameters of the theory. In the case of SU(2)
symmetry in the spin subspace the interaction term Eq.\ (\ref{Hintz})
yields a rescaled  $K_{\mu}\to K_{\mu}^*=1$.

It must be emphasized that the one-dimensional idealization of the
magnetic subsystem albeit practical has to be used with caution. As
discussed in detail in Ref.\ \onlinecite{Wern02a} in the Fermi liquid
regime for $T < 25\ {\rm K} \sim \overline{V}_{\rm m}$ the magnetic
correlations are two-dimensional but with corrections imposed by the
closeness to one-dimensionality. Here the one-dimensional expressions
will be applied to make use of results from the literature available
for those systems but a discussion of the expected applicability to the
realistic case including the two-dimensional coupling is given at the
same time (Secs.\ \ref{sectionsG} and \ref{sectionPCE}).

The back and umklapp scattering term in the bosonized Hamiltonian is
\begin{eqnarray}\label{Hintz}
H_{\rm int}&=& \frac{U_0}{(2\pi a)^2} \int_{} d^2\ovr
    \cos \sqrt{4\pi}\phi_{\rm s}(\obr)\, 
                     \cos \sqrt{4\pi}\phi_{\rm sf}(\obr)
\nonumber\\&+& \frac{1}{(2\pi a)^2} \int_{} d^2\ovr\
\cos \big[\sqrt{4\pi}\phi_{\rm f}(\obr) - 2\sqrt{2} k_{\rm F} \oy\big]
\nonumber\\[-0ex]&&\hspace{4ex}\times
 \Big(U_1\cos \sqrt{4\pi}\phi_{\rm s}(\obr) + 
                    U_2\cos \sqrt{4\pi}\phi_{\rm sf}(\obr)\Big)
\nonumber\\&+& \frac{1}{(2\pi a)^2} \int_{} d^2\ovr\
\cos \big[\sqrt{4\pi}\phi_{\rho}(\obr) - 2\sqrt{2} k_{\rm F} \ox\big]
\nonumber\\[-0ex]&&\hspace{4ex}\times
 \Big(U_2\cos \sqrt{4\pi}\phi_{\rm s}(\obr) + 
                    U_1\cos \sqrt{4\pi}\phi_{\rm sf}(\obr)\Big).
\nonumber\\[-0ex]&&
\end{eqnarray}
The limit $a\to 0$ and $L\to \infty$ is understood. Note that the
symmetry of the superconducting saddle point (Sec.\
\ref{sectionOrder}) deduced from the interaction term Eq.\
(\ref{Hintz}) is manifestly independent of corrections to the
bosonized approach since the energy scale is at least an order of
magnitude larger than the correction terms, i.e., $U_{0,1,2} \gg
t_{i,h,\perp,z} > \overline{V}_{\rm m}$ (c.f.\ Ref.\
\onlinecite{Wern02a}).   

Equations (\ref{Hc(m)}) through (\ref{Hintz}) are explicitly invariant 
under the symmetry operations of the tetragonal lattice.


\subsection{Bosonized pair operators}\label{sectionBose}

At $\bq=0$ the pair operators defined in Eqs.\
(\ref{efpair}) -- (\ref{mfs1pair})  are local, i.e.,
$P_{a,b}(\bq=0)=\sum_\bl P_{a,b}(\bl)$. The correlations in the
$P_{0,\pm}$ channel have been shown to be negligible with respect to
the mixed flavor channels in Sec.\ \ref{sectionnormal}. In the
continuum representation the pair operators can be given in terms of
the Bose fields as introduced in Ref.\ \onlinecite{Wern02a}. The
abbreviations  $\cos\sqrt{\pi}\phi_\mu(\obr) = c\phi_\mu$,
$\sin\sqrt{\pi}\phi_\mu(\obr) = s\phi_\mu$,
$\cos\sqrt{\pi}\theta_\mu(\obr) = c\theta_\mu$,
$\sin\sqrt{\pi}\theta_\mu(\obr) = s\theta_\mu$, as well as $\ophi_{\rm
f}=\phi_{\rm f} - \sqrt{2/\pi}\, k_{\rm F} \, \oy$ and
$\ophi_{\rho}=\phi_{\rho} - \sqrt{2/\pi} \,  k_{\rm F} \, \ox$ are
introduced for clarity. The pair operator components Eqs.\
(\ref{mfs0pair}) and (\ref{mfs1pair}) then are 
\begin{eqnarray}\label{pair1+}
P_{1,+}(\obr) &\propto& \frac{2 {\rm e}^{-i\sqrt{\pi}\theta_{\rho}}}{\pi a}
\Big[
c\theta_{\rm sf}\,c\ophi_{\rm f}\,c\phi_{\rm s}
                  + i\, s\theta_{\rm sf}\,s\ophi_{\rm f}\,s\phi_{\rm s}
\nonumber\\&&\hspace{8ex}
+\ c\theta_{\rm sf}\,c\ophi_{\rho}\,c\phi_{\rm sf}
                  + i\, s\theta_{\rm sf}\,s\ophi_{\rho}\,s\phi_{\rm sf}\Big]
\\\label{pair1-}
P_{1,-}(\obr) &\propto& \frac{-2 {\rm e}^{-i\sqrt{\pi}\theta_{\rho}}}{\pi a}
\Big[
c\theta_{\rm sf}\,s\ophi_{\rm f}\,s\phi_{\rm s}
                  + i\, s\theta_{\rm sf}\,c\ophi_{\rm f}\,c\phi_{\rm s}
\nonumber\\&&\hspace{8ex}
+\ c\theta_{\rm sf}\,s\ophi_{\rho}\,s\phi_{\rm sf}
                  + i\, s\theta_{\rm sf}\,c\ophi_{\rho}\,c\phi_{\rm sf}\Big]
\end{eqnarray}
\begin{eqnarray}\label{pair2+}
P_{2,+}(\obr) &\propto& \frac{2 {\rm e}^{-i\sqrt{\pi}\theta_{\rho}}}{\pi a}
\Big[
c\theta_{\rm s}\,c\ophi_{\rm f}\,c\phi_{\rm sf}
                  + i\, s\theta_{\rm s}\,s\ophi_{\rm f}\,s\phi_{\rm sf}
\nonumber\\&&\hspace{8ex}
+\ c\theta_{\rm s}\,c\ophi_{\rho}\,c\phi_{\rm s}
                  + i\, s\theta_{\rm s}\,s\ophi_{\rho}\,s\phi_{\rm s}\Big]
\\\label{pair2-}
P_{2,-}(\obr) &\propto& \frac{-2 {\rm e}^{-i\sqrt{\pi}\theta_{\rho}}}{\pi a}
\Big[
c\theta_{\rm s}\,s\ophi_{\rm f}\,s\phi_{\rm sf}
                  + i\, s\theta_{\rm s}\,c\ophi_{\rm f}\,c\phi_{\rm sf}
\nonumber\\&&\hspace{10ex}
+\ c\theta_{\rm s}\,s\ophi_{\rho}\,s\phi_{\rm s}
                  + i\, s\theta_{\rm s}\,c\ophi_{\rho}\,c\phi_{\rm s}\Big]
\end{eqnarray}
The limit $a\to 0$ is understood. The Klein factors have not been
plotted here for lucidity but are of crucial importance in the present
model of multiple electron species when calculating correlation
functions. They do not intervene with the symmetry arguments for the
static expectation values discussed below.

A similar analysis of the spin density operators 
\begin{equation}
S^\mu(\bl) = \sum_{\sigma,\sigma'}\ \sum_{\nu=x,y}
         \sigma^\mu_{\sigma,\sigma'}\
c^{\dagger}_{{\bl},\nu,\sigma}\, 
                     c^{\phantom{\dagger}}_{{\bl},\nu,\sigma'} 
\end{equation}
reveals that the out-of-plane component $S^z(\bl) \sim f(\phi_{\rm s},
\phi_{\rm sf})$ depends on the Bose fields while the in-plane
components $S^x(\bl) \sim S^y(\bl) \sim f(\theta_{\rm s}, \theta_{\rm
sf})$ depend on the dual fields. The $\sigma^\mu_{\sigma,\sigma'}$ are
the Pauli matrices. Consequently $K^*_{\rm s} < K^*_{\rm sf} \le 1$ is
consistent both with the observed magnetic easy plane configuration in
Sr$_2$RuO$_4$ (Sec.\ \ref{sectionOP} and Ref.\ \onlinecite{Wern02c})
{\em and} with $\chi^{(P)}_{\rbx{1,b}}(\bq,\omega) <
\chi^{(P)}_{\rbx{2,b}}(\bq,\omega)$. $\chi^{(P)}_{\rbx{a,b}}$ was
defined in Eq.\ (\ref{chiPabnull}). The bosonized model is consistent 
with the superconducting phase transition in the $\{P_{2,+},P_{2,-}\}$
sector. 

As discussed in Sec.\ II C of Ref.\ \onlinecite{Wern02a} the bosonized 
approach neglects the exchange terms of Hund's rule coupling. This
leads to an underestimation of the $S^z = 0$ spin-triplet component
$P_{1,+}$. The $\mu$SR measurements\cite{LFK+98} discussed in Sec.\
\ref{sectionOP} and the upper critical fields (Ref.\
\onlinecite{Wern02c}) suggest that $\langle P_{1,+} \rangle 
\approx 0.8 \langle P_{2,+} \rangle \approx 0.8 \langle P_{2,-}
\rangle$. Since this is close to the isotropic limit
$\chi^{(P)}_{\rbx{1,+}}(\bq,\omega) \approx
\chi^{(P)}_{\rbx{2,\pm}}(\bq,\omega)$ will be assumed while
$\chi^{(P)}_{\rbx{1,-}}(\bq,\omega) < 
\chi^{(P)}_{\rbx{1,+}}(\bq,\omega)$ and consequently the spin-singlet
component does not contribute, i.e., $\langle P_{1,-} \rangle =
0$. The instability occurs in the spin-triplet channel consistent with
the results in Sec.\ \ref{sectionnormal}.

\subsection{Order parameter}\label{sectionOrder}

From Eqs.\ (\ref{pair1+}), (\ref{pair2+}) and (\ref{pair2-}) follows
that the finite value of the order parameter $\langle P_{2,\pm}
\rangle, \langle P_{1,+} \rangle \neq 0$ implies $\langle
\exp(-i\sqrt{\pi}\theta_\rho) \rangle \neq 0$. This is
incompatible\cite{CT02} with finite expectation values of operators
containing the dual field and thus terms $\sim \cos\sqrt{\pi} \,
\ophi_\rho$ and $\sim \sin\sqrt{\pi} \, \ophi_\rho$ are discarded. 

The interaction that led to the derivation of $H_{\rm mf}$ in Eq.\
(\ref{Hmf}) is attractive. A Tomonaga-Luttinger model with attractive   
interaction scales to strong coupling under a renormalization group
analysis and has a gapped excitation spectrum.\cite{Emer79,SCP98} Its
thermodynamic properties can be modeled via the sine-Gordon
model.\cite{LE74,Luth77,Tsve95,GNT98} The free energy is lowered
proportional to the square of the excitation gap.\cite{Baxt82,LZ97}
The energy can be minimized by maximizing the gap. This suggests
the minimization of the repulsive components of the interaction term 
$H_{\rm int}$ through $
\langle \cos\sqrt{4\pi}\phi_{\rm s} \rangle =
\langle \cos\sqrt{4\pi}\phi_{\rm sf} \rangle =
\langle \cos\sqrt{4\pi}\,\ophi_{\rm f} \rangle =
\langle \cos\sqrt{4\pi}\,\ophi_{\rho} \rangle = 0
$. It is then
consistent to assume that these operators are irrelevant and
consequently the contributions $\sim \cos \sqrt{\pi} \phi_{\rm s}$,
$\sim \cos \sqrt{\pi} \phi_{\rm sf}$, $\sim \cos \sqrt{\pi}\,
\ophi_{\rm f}$, and $\sim \cos \sqrt{\pi}\, \ophi_{\rho}$ in the
operators $P_{2,\pm}$ and $P_{1,+}$ can be neglected. 

One then obtains a gap in all channels $\mu\in\{{\rm c}, {\rm s}, {\rm 
sf}\}$\cite{gapquote} with possible finite expectation values for 
$\langle \sin\sqrt{\pi}\phi_{\rm sf} \rangle$, 
$\langle \sin\sqrt{\pi}\theta_{\rm s} \rangle$, 
$\langle \cos\sqrt{\pi}\theta_{\rm s} \rangle$, and
$\langle \sin\sqrt{\pi}\ophi_{\rm f} \rangle$ as will now be discussed
in detail.

The system is invariant under a global phase shift in the charge
sector. The fluctuations of the charge phase are referred to as the
Anderson-Goldstone collective mode.\cite{Ande58} They can be
parameterized by introducing the unit vector 
\begin{equation}\label{Omegarho}
\bOmega_\rho = 
\left({-i\, \sin\sqrt{\pi}\theta_\rho \atop 
       \cos\sqrt{\pi}\theta_\rho} \right)\,. 
\end{equation}
The Anderson-Goldstone collective mode results from a broken Gauge
symmetry and in the presence of long range Coulomb interaction acquires
a gap of the order of the plasma frequency through the Anderson-Higgs
mechanism.\cite{Ande58,Schu95} The large gap allows for neglecting the
charge phase fluctuations in the discussions of the low energy
excitations that follow (Sec.\ \ref{sectionfluct}).

Correspondingly, fluctuations of $P_{2,+}$ and $P_{2,-}$ can be
parameterized as phase fluctuations in the spin channel via
$\theta_{\rm s}$ which takes the role of a azimuthal angle of the
magnetic moment of the Cooper pairs. The third spin-triplet component
$P_{1,+}$ can be included via a polar angle $\theta_z$.
\begin{equation}\label{Omegas}
\bOmega_{\rm s} = 
\left(\begin{array}{c}
     -i\ \sin\sqrt{\pi}\theta_{\rm s} \sin \theta_{z} \cr 
       \cos\sqrt{\pi}\theta_{\rm s} \sin \theta_{z} \cr 
       \cos \theta_{z} \end{array}\right)
\end{equation}
The SO(3) vector $\bOmega_{\rm s}$ describes the three spin-triplet
components introduced in Sec.\ \ref{sectionnormal}. 

The order parameter expectation values $\langle P_{2,\pm} \rangle,
\langle P_{1,+} \rangle \sim \langle \sin \sqrt{\pi} \ophi_{\rm f}
\rangle = \langle \sin (\sqrt{\pi} \phi_{\rm f} - \sqrt{2}\, \oy)
\rangle$ break the invariance of the model under reflection $y\to
-y$. This in an artifact of the mean-field decoupling Eq.\
(\ref{mfdecoupling}) which yields only one of two equivalent saddle
points. In the absence of external fields, temperature gradients, or
strain the symmetry must be restored by including another
two-component vector\cite{Sigr00,Agte01}   
\begin{equation}\label{Omegaf}
\bOmega_{\rm f} = 
\left( \sin (\sqrt{\pi} \phi_{\rm f} - \sqrt{2}\, \oy) \atop 
       \sin (\sqrt{\pi} \phi_{\rm f} - \sqrt{2}\, \ox) \right)\,. 
\end{equation}
Equation (\ref{Omegaf}) includes a $p$-wave component of the pair
operator. This becomes apparent in the rewritten form  
\begin{equation}\label{pcomp}
\bOmega_{\rm f} = 
\left( \cos \sqrt{2}\oy\, \sin \sqrt{\pi} \phi_{\rm f} \atop 
       \cos \sqrt{2}\ox\, \sin \sqrt{\pi} \phi_{\rm f} \right) 
-
\left( \sin \sqrt{2}\oy\, \cos \sqrt{\pi} \phi_{\rm f} \atop 
       \sin \sqrt{2}\ox\, \cos \sqrt{\pi} \phi_{\rm f} \right)  . 
\end{equation}
Since the component $\Omega_{{\rm f},x}$ is oscillatory along $\oy$
the pair correlations are stronger along $\ox$ and vice versa. The
existence of such two order parameter components that couple
differently to magnetic fields through a slight spatial anisotropy has
been implied experimentally via upper critical
field\cite{YAM+01,Agte01,Wern02c} and thermal
conductivity\cite{ITY+01} measurements as discussed closer in Sec.\
\ref{sectionsymm}. The fluctuations of $\bOmega_{\rm f}$ account for
the observed strain anomalies\cite{OSM+02} in the superconducting
phase (Sec.\ \ref{sectionfluct}).

Introducing the normalized average 
\begin{equation}\label{Omegaaver}
\obOmega_{\mu} = \langle \bOmega_\mu \rangle / |\langle \bOmega_\mu \rangle| 
\end{equation} 
$|\obOmega_\mu|^2=1$ is satisfied. Denoting the order parameter
amplitude
\begin{equation}\label{Paverage}
\oP = |\langle P \rangle| = \frac{2}{\pi}\, 
    |\langle\sin\sqrt{\pi}\phi_{\rm sf}\rangle|\                  
     |\langle \bOmega_\rho \rangle|\ 
|\langle \bOmega_{\rm s} \rangle|\ |\langle \bOmega_{\rm f} \rangle|
\end{equation}
the mean-field Hamiltonian Eq.\ (\ref{Hmf}) becomes in its continuum
representation 
\begin{eqnarray}\label{Hmfeff}
H_{\rm mf} &=& V_{0} \int_{-L}^L\! \frac{d^2\ovr}{a^2} \bigg[
-  \oP^2
\nonumber\\&&\hspace{-2ex}+\
\frac{4\oP}{\pi}
\sin\sqrt{\pi}\phi_{\rm sf}
        \left(\obOmega_{\rho}^*\, \bOmega_{\rho}\right)\!
           \left(\obOmega_{\rm s}^*\, \bOmega_{\rm s}\right)\!
              \left(\obOmega_{\rm f}^*\, \bOmega_{\rm f}\right)\!\bigg].\!
\end{eqnarray}

Together with the in-plane contributions Eqs.\ (\ref{Hc(m)}) and
(\ref{effHmu}) the Hamiltonian in the superconducting phase is given
by   
\begin{equation}\label{Hsc}
H_{\rm sc} = 
       H_{\rm c} + H_{\rm s} + H_{\rm sf} + H_{\rm mf}\,.
\end{equation}


\subsection{Magnetic and temperature fields}\label{sectionfields}

Magnetic fields need to be included in the microscopic model by
introducing a vector potential in the momentum operator.\cite{Tink96}
A detailed study within the model proposed here is rather involved and
is left for future studies. One can use the notion of spin-charge
separation thought to discuss some qualitative implications of an
applied magnetic field. The coupling of an applied magnetic field
$\bh$ to the magnetic degrees of freedom of the order parameter then
is simply given by 
\begin{equation}\label{Hh}
H_h = -  \int_{-L}^L d^2\ovr\ \bh\, \obOmega_{\rm s}
\end{equation}
This term further reduces the symmetry in the spin channel, e.g.,
$\obOmega_{\rm s} = (\sqrt{2}^{-1}, \sqrt{2}^{-1}, 0)^\dagger$ if the
quantization axis is chosen along the applied field. The ground state
order parameter then carries spin one along the direction of the
magnetic field. 

The study of the coupling of an external magnetic field to the
anisotropic components of the flavor vector $\obOmega_{\rm f}$ is more
involved. Related two-component superconductors have been studied via
a Landau Ginzburg analysis\cite{Sigr00,Agte01} albeit based on $p$-wave
symmetries without the notion of gapless fluctuations addressed in
Sec.\ \ref{sectionfluct}. The general physical result though is that
the magnetic field couples strongest to the order parameter component
with dominant superconducting correlations perpendicular to the
field. If the applied field has a component in the $x$-$y$ plane, the
coupling to the components of $\bOmega_{\rm f}$ will be
inequivalent. A field along $\ox$ will couple stronger to the order
parameter component $\Omega_{{\rm f},y}$ since it has dominant pair
correlations along $\oy$.  

The application of a temperature gradient in the $x$-$y$ plane breaks
the reflection symmetry of the lattice and yields inequivalent
components of $\obOmega_{\rm f}$. The order parameter symmetry is
reduced to a two-fold axis\cite{ITY+01} perpendicular to the $x$-$y$
plane as opposed to a four-fold axis in the absence of the temperature
gradient. Similar effects are expected through strain\cite{OSM+02} and
at surfaces or grain\cite{SM01} boundaries.

\subsection{Mapping onto sine-Gordon models}\label{sectionsG}

The mean-field contribution Eq.\ (\ref{Hmfeff}) couples all the
channels $\mu \in \{\rho, {\rm s}, {\rm f}, {\rm sf}\}$. Since the 
lowest bound state and thus the gap energy is correctly reproduced by
a perturbative treatment even in one-dimensional interacting
systems\cite{LZ97} one can attempt a mean-field like decoupling valid
at least for small values of the fields $\phi_\mu$. To this end all
terms $\sim\sin\sqrt{\pi}\phi_\mu$ and $\sim\sin\sqrt{\pi}\theta_\mu$
are transformed to $\cos$ terms via the sliding transformations
$\phi_\mu \to \phi_\mu - \frac{\pi}{4}$ and $\theta_\mu \to \theta_\mu
- \frac{\pi}{4}$. Expanding the $\cos$ terms to second order in the
fields decouples $H_{\rm mf,eff}$ in the different channels
$\mu$. Rewriting the resulting contributions $H_{{\rm mf},\mu}$ in
terms of $\cos$ and reversing the sliding transformations $\theta_\mu
\to \theta_\mu + \frac{\pi}{4}$ where necessary yields the desired
result (up to a constant).

In order to obtain the Lagrangian for the Hamiltonians in Eq.\
(\ref{effHmu}) Euclidean time dependence of the fields is introduced, 
the Lagrange transformation $\Pi_\mu(\obr,\tau) = \frac{i}{K^*_\mu}
\dot{\phi}_\mu(\obr,\tau)$ is performed, and the fields are rescaled
as $\varphi_\mu = (K^*_\mu)^{-1/2}\, \phi_\mu$. The effective,
one-dimensional action for the spin-flavor channel is obtained as  
\begin{equation}\label{Seffsf}
S_{\rm sf}^{\rm eff}\!=\!\frac{v_{\rm sf} L}{2}\!\!
      \int\limits_{-L}^L\!\! d\oy\! \int\limits_{0}^\beta\!\! d\tau 
    \left[\dot{\varphi}^2_{\rm sf} + 
          (\partial_\oy \varphi_{\rm sf})^2 \! + 
  \oP M_{\rm sf} \cos B_{\rm sf} \varphi_{\rm sf}\right] 
\end{equation}
with $B_{\rm sf} = \sqrt{\pi K^*_{\rm sf}}$ and inverse temperature
$\beta=T^{-1}$. Equation (\ref{Seffsf}) is the quantum sine-Gordon
action with the interaction parameter $M_{\mu} = \frac{8 V_{0}}{
v_\mu \pi a^2}$.

The transformation to the dual fields $\Pi_\mu = -\partial_\ox
\theta_\mu$ and rescaling as $\vartheta_\mu = (K^*_\mu)^{1/2}\,
\theta_\mu$ yields the effective, one-dimensional spin-channel action   
\begin{equation}\label{Seffs}
S_{\rm s}^{\rm eff}=\frac{v_{\rm s} L}{2}\!\!
      \int\limits_{-L}^L\!\! d\ox \int\limits_{0}^\beta\!\! d\tau  
    \left[\dot{\vartheta}^2_{\rm s} + (\partial_\ox
                                \vartheta_{\rm s})^2 +  
\oP M_{\rm s}\, \obOmega'_{\rm s}\, \bOmega'_{\rm s} \right]
\end{equation}
for correspondingly adapted vectors $\obOmega'_{\rm s}$ and
$\bOmega'_{\rm s}$. The action in Eq.\ (\ref{Seffs}) is invariant
under a rotation of $\obOmega'_{\rm s}$. Transversal fluctuations of
$\bOmega'_{\rm s}$ with respect to $\obOmega'_{\rm s}$ give rise to
Goldstone modes which can be treated separately as will be done in
Sec.\ \ref{sectionfluct} and Appendix A. To study thermodynamic 
properties as implied by the sine-Gordon action one can consequently
use the simplified expression with $\obOmega'_{\rm s}\, \bOmega'_{\rm
  s} \to \cos B_{\rm s} \vartheta_{\rm s}$ with $B_{\rm s} =
\sqrt{\pi}/\sqrt{K^*_{\rm s}}$.\cite{SineGordonSquote}

For $T\to 0$ and $L\to\infty$ the sine-Gordon actions Eqs.\
(\ref{Seffsf}) and (\ref{Seffs}) are equivalent to the charge
zero-sector of the massive Thirring model if $B_\mu <
8\pi$.\cite{Cole75,LZ97} For both channels to be in the gapped regime  
requires $K^*_{\rm sf}<8$ and $K^*_{\rm s}>1/8$. Unless $K^*_{\rm
s} < 1$ is very strongly renormalized the conditions on $K^*_\mu$ are
satisfied.

For $T\to 0$ and $L\to\infty$ the actions Eqs.\ (\ref{Seffsf}) and
(\ref{Seffs}) have a gap of\cite{LZ97}
\begin{equation}\label{DeltaGap} 
\Delta_\mu \sim \left[\oP\ a^2 M_\mu
           \right]^{\frac{1}{2-(B_\mu^2)/(4\pi)}}
\end{equation}
to the lowest bound state above the ground state.

Similarly the action for the charge channel can be derived.
\begin{eqnarray}\label{Sc}
S_{\rm c} &=&   
\frac{1}{2} \int_{-L}^L\!\! d^2\ovr \int_{0}^\beta\!\! d\tau\,
    \bigg\{\vF \left(\dot{\phi}^2_{\rho} + \dot{\phi}^2_{\rm f}\right) 
\nonumber\\&&\hspace{2ex}
 +\ V_{\rm c} \left[\partial_{\ox} \phi_{\rho}
    + \partial_{\oy} \phi_{\rm f}\right]^2
 + \overline{V}_{\rm\!\!c} 
                   \left[\partial_{\oy} \phi_{\rho}
    + \partial_{\ox} \phi_{\rm f}\right]^2
\nonumber\\&&\hspace{10ex}
 +\ \oP\ \frac{4 V_{0}}{\pi a^2}
        \left(\obOmega_{\rho}^*\, \bOmega_{\rho}\right)
              \left(\obOmega_{\rm f}^*\, \bOmega_{\rm f}\right)
\bigg\},
\end{eqnarray}
The action Eq.\ (\ref{Sc}) can be transformed to both dual fields
$\theta_\rho$ and $\theta_{\rm f}$ but a mixed representation in
$\theta_\rho$ and $\phi_{\rm f}$ as desirable is not possible.

No detailed theoretical results are known for the two-dimensional
two-component sine-Gordon model defined by Eq.\ (\ref{Sc}). In the
gapped phase relevant here the difference between the two-dimensional
correlations and the one-dimensional ones are less severe (Sec.\
\ref{sectionPCE}) then in the absence of a gap. It is thus permissible
to use the theoretical results given in the literature for the
one-dimensional actions Eqs.\ (\ref{Seffsf}) and (\ref{Seffs}) and
anticipating corrections from the two-dimensional character of Eq.\
(\ref{Sc}) and the two-dimensional corrections\cite{Wern02a} to Eqs.\
(\ref{Seffsf}) and (\ref{Seffs}).


\subsection{Fluctuations}\label{sectionfluct}

The description of Goldstone modes in symmetry broken phases via
non-linear sigma models is a standard problem in text book
literature.\cite{Frad91,Tsve95,GNT98} The description of the charge
phase fluctuations including the Anderson-Higgs mechanism leading to
the charge-phase gap has been derived by Schulz\cite{Schu95} for the 
case of the attractive two-dimensional Hubbard model. Goldstone modes
of the magnetically ordered phase of the repulsive two-dimensional
Hubbard model have been discussed by Schulz\cite{Schu95,Schu90b} and
Weng\cite{WTL91} {\em et al}. Most importantly, the present approach
is a two-dimensional analog of superfluid
$^3$He-$A$.\cite{VK83,Volo92} The conclusive result is that for
temperatures sufficiently smaller than the electronic excitation gap
$T \ll \Delta$ and at sufficiently small energies $\omega \ll \Delta$
the gapless fluctuations of the internal degrees of freedom of the
order parameter associated with continuously broken symmetries are
well described by non-linear sigma models (Appendix A). 

In the present approach the gapless Goldstone modes correspond to the
broken local spin rotational invariance described by the SO(3) vector
$\bOmega_{\rm s}$ and the degenerate flavor saddle points described
by the SO(2) vector $\bOmega_{\rm f}$ with local constraints
$|\bOmega_{\mu}(\br,\tau)|^2 = 1$. 

At sufficiently small temperatures and energy scales $T,\omega \ll
\Delta$ the stiffness of the non-linear sigma model describing the
angular fluctuations has been determined\cite{Tsve95,GNT98} to scale
with the square of the order parameter amplitude $\sim \oP^2$. The
analogy to magnetic excitations in the ordered Hubbard model suggest
corrections\cite{WTL91,Schu95} that depend on the specific
correlations of the system. The resulting non-linear sigma model in
Euclidean space (Appendix A) is\cite{Schu95} 
\begin{equation}\label{Snls}
S_\Omega = 
  \oP^2 \int\limits_{-L}^L\! d^2r 
                 \int\limits_0^\beta d\tau 
\!\sum_{\nu,\mu}\!A_{\nu}\Bigg[
\frac{(\dot{\obOmega}_{\mu})^2}{(a v_{\nu})^{2}}
+ ( \partial_\nu \obOmega_{\mu})^2
\Bigg] ,
\end{equation}
the stiffness parameters are   
\begin{equation}\label{stiffness}
A_{\nu} = \frac{\partial^2}{\partial q^2_\nu} \left(V_{\bq}^2\
    \chi^{(P)}_{\rbx{2,\pm}}(\bq,0)\right)\Big|_{\bq=0}
\end{equation}
with the excitation velocities
\begin{equation}\label{vOmega}
\frac{1}{(v_{\nu})^{2}}= \frac{V_{0}^2}{A_{\nu}}\,
\frac{\partial^2}{\partial \omega^2}
    \chi^{(P)}_{\rbx{2,\pm}}(0,\omega) \Big|_{\omega=0}\,.
\end{equation}
Here the magnetic channel is treated isotropically. The pair
correlation functions depend on $q_\ox,q_\oy \sim q_x\pm
q_y$. Consequently $A_x=A_y$ reflecting the symmetry of the 
lattice. The stiffness perpendicular to the $xy$ plane is negligible
since the out-of-plane correlations are two orders of magnitude
smaller as indicated by the resistivity,\cite{MYH+97} thermal
conductivity,\cite{TSN+01} and critical magnetic field\cite{ANM99}
measurements. Within the mean-field approach $A_z \equiv 0$. 
Note that Eqs.\ (\ref{Snls}) through (\ref{vOmega}) can also be
obtained by replacing $X^2 \to \oP^2$ in the amplitude fluctuation
expression Eq.\ (\ref{SOmega}) in Appendix B.

The crucial difference of the present model with respect to BCS theory
is the presence of gapless modes in the ordered phase. Unlike in BCS
the excitations of the internal degrees of freedom of the order
parameter, namely the SO(3) spin-triplet components and the SO(2)
flavor components do not necessarily acquire a gap (Appendix B2). The
analysis of a related model\cite{Baba02} and the
analogy\cite{VK83,Volo92} to $^3$He-$A$ suggests that the magnetic
excitations remain gapless and have linear 
dispersions.\cite{Spinspectrumquote} Similar
arguments\cite{VK83,Volo92} hold for the flavor degrees of
freedom.\cite{Flavorspectrumquote} 

Note that even in the case of a large gap $\Omega_A \ge \Delta$ of the
magnetic fluctuations $\bOmega_{\rm s}$ which could in principle
induced by the spin-flavor coupling terms in Eq.\ (\ref{Hintz}) the
physical implications of the present approach remain
unaltered because of the ungapped SO(2) flavor mode\cite{Volo92}
$\bOmega_{\rm f}$. The degeneracy of the components of $\bOmega_{\rm
  f}$ are prerequisite for the presence of a gapless flavor mode. At
this point it should be stressed that the tetragonal structure of
Sr$_2$RuO$_4$ has been found to be very stable (c.f.\ Ref.\
\onlinecite{BRN+98}). Since $\Omega_{{\rm f},x}$ and $\Omega_{{\rm
    f},y}$ are related via a symmetry operation under which  the
tetragonal point group of the lattice is invariant it can be concluded
that the degeneracy of the two components is also very stable, at
least in the bulk material. For a discussion see Sec.\
(\ref{sectionDisc}).

Experimental evidence of the existence of the flavor mode stems from
ultrasonic measurements in Ref.\ \onlinecite{OSM+02} that probe the
strain dependence of the superconductivity in Sr$_2$RuO$_4$. The
in-plane anisotropy of the components $\Omega_{{\rm f},x}$ and
$\Omega_{{\rm f},y}$ leads to a coupling of the superconducting order
parameter to the strain and thus accounts for the observed
effects. The coupling to out-of-plane shear strain components is much
weaker. The gapless excitations yield a reduction of the elastic
constants in the superconducting phase. The strain breaks the symmetry
between the two components which accounts for the anomalies observed
near the phase transition. Also, the observed anomalous absorption in
cyclotron resonance experiments\cite{RES+02} might be attributed to the
coupling to the $\bOmega_{\rm f}$ gapless mode.

\subsubsection{Estimates for the pair correlation function}\label{sectionPCE}

On the mean-field level one expects that the pair correlation
functions that determine the the stiffness and
excitation velocities through Eqs.\ (\ref{stiffness}) and
(\ref{vOmega}) have the form\cite{Blat64,Tink96}
\begin{equation}\label{chiBCS}
\chi^{(P)}_{\rbx{\rm BCS}}(\bq,\omega) = \frac{- \Delta f(\bq)}
    {\omega^2 - \ov_{\rm eff}^2 \bq^2 - \Delta^2}\,,
\end{equation}
where $\Delta$ is the single particle energy gap, $\ov_{\rm eff}$ is
the effective velocity of the collective excitations associated with
the pair operators, and $f(\bq)$ is a function that weakly depends on
$\bq$. Note that in the static limit and with a pair correlation
length $\xi_P$ related to the superconducting gap\cite{Blat64} as
$\xi_P \propto \ov_{\rm eff}/\Delta$ the pair correlation function has
Lorentzian form $\chi^{(P)} (\bq\to0, \omega=0) \sim \xi_P^{-1} (|a
\bq|^2 + \xi_P^{-2})^{-1}$. Equation (\ref{chiBCS}) is  thus valid on
a rather general basis.

Since one expects that the quasi one-dimensional correlations exhibit
the most singular behavior, and since the normal phase magnetic
properties of the system are dominated by quasi one-dimensional
correlations,\cite{Wern02a} it is necessary to investigate the
possible impact of the one-dimensional subsystem on the stiffness and
excitation velocities through Eqs.\ (\ref{stiffness}) and
(\ref{vOmega}). 

First of all it must be stated that the scale invariance of the
magnetic correlations breaks down\cite{BSB+02} in the Fermi liquid
regime for $T < 25$ K indicating that the system turns two-dimensional
at low temperatures.\cite{Wern02a} Furthermore, the pair correlation
functions factorize into spin, spin-flavor, charge, and flavor
contributions in the bosonized representation [c.f., Eqs.\
(\ref{pair1+}) through (\ref{pair2-})]. In the superconducting phase
the correlation functions have to be evaluated with respect to the
actions Eqs.\ (\ref{Seffsf}), (\ref{Seffs}), and (\ref{Sc}) where all
channels are gapped. Consequently in the superconducting phase the
divergent correlations of the quasi one-dimensional magnetic subsystem
are suppressed and $\chi^{(P)}_{\rbx{2,\pm}}(\bq,\omega)$ is analytic
in $\bq$ and $\omega$. The contributions from the charge and flavor
channels are two dimensional anyway [Eq.\ (\ref{Sc})].

To make these reflections more transparent consider that for $T >
T_{\rm c}$ the spin and spin-flavor channel correlation functions are
approximately determined by the Hamiltonian Eq.\ (\ref{effHmu}). They
have the standard functional dependence in conformal field
theory.\cite{LP75b,Tsve95} Density and pair correlation functions are
connected via substituting $|2k_{\rm F} - q_\onu| \to q_\onu$ with
Fermi wavenumber $k_{\rm F}$ and an appropriate adoption of the
scaling dimension.\cite{Emer79}  
Dimerized spin chains are described by the action Eq.\ (\ref{Seffsf})
with\cite{US96} $B^2_\mu=2\pi$ and the density correlation functions
have been given in Ref.\ \onlinecite{ETD97}. The value of
$B^2_\mu=2\pi$ is not a singular  point in the analysis\cite{LZ97} at
$T=0$ and similar behavior for other values of $B^2_\mu$ is expected,
especially at finite temperatures. This leads to the
estimate\cite{ETD97} 
\begin{equation}\label{chi1D}
\chi^{(P)}_{\rm 1D}(q_\onu,\omega) = \frac{A_0\ 
       \cos^2(a q_\onu/2)}
    {\omega^2 - \ov_{\rm eff}^2 \sin^2 (a q_\onu) - \Delta^2}\,,
\end{equation}
where $\Delta$ is the energy gap to the lowest bound state. Comparing
with Eq.\ (\ref{chiBCS}) one consequently finds that in the gapped phase
$\chi^{(P)}_{\rm 1D} (q_\onu,\omega) \equiv \chi^{(P)}_{\rbx{\rm BCS}}
(q_\onu,\omega)$ with $-\Delta f(q_\onu) = A_0\ \cos^2(a
q_\onu/2)$. Moreover, $\chi^{(P)}_{\rm 1D}(q_\onu,0)$ is analytic in
$q_\onu = 0$. This result is not surprising in the light of the relevant
three-dimensional couplings that have to be present in order to drive
any quasi one-dimensional system into the ordered phase.\cite{ETD97} 

In Appendix B1 it is shown that the analytical properties of
the pair correlation functions $\chi^{(P)}_{\rbx{2,\pm}}$ are related
to those of $\chi^{(P)}_{\rbx{\rm BCS}}$ [c.f.\ Eq.\
(\ref{chiidentify})]. Given the analyticity of the one-dimensional
contribution Eq.\ (\ref{chi1D}), the coupling terms that render the
magnetic system two-dimensional below\cite{Wern02a} $T < 25$ K, and
the two-dimensionality of the charge contribution as shown in Eq.\
(\ref{Sc}) it can be concluded that the $A_\nu$ as defined in Eq.\
(\ref{stiffness}) are dominantly isotropic in the $q_x$-$q_y$ plane
for $T \ll \ov_{\rm eff}$ and $q_\nu \ll a^{-1}$ up to quadratic order
in $q_\nu$. 

From the geometry of the upper critical magnetic fields discussed in
Ref.\ \onlinecite{Wern02c} it can be concluded that there is an
in-plane anisotropy in the order parameter of $\sim 7$\% which should
be reflected also in $\chi^{(P)}_{\rbx{2,\pm}}$. Most importantly, the
non-analyticity along the diagonals of the basal plane\cite{Wern02c}
suggests the presence of a small term $\sim |q_x q_y|$ in the
expansion of $\chi^{(P)}_{\rbx{2,\pm}}$. Considering the small energy
scale set by the anisotropic component of the upper critical magnetic
field $H_{{\rm c},a}(T\to 0) \approx 0.05\ {\rm T} \sim 0.04$ K the
effect should be negligible here for $T \ge 0.1$ K. The expected
significant next higher order anisotropic correction terms are $\sim
q_x^2q_y^2$.

\subsubsection{Effective Saddle point}\label{sectionSaddle}

The amplitude fluctuations of the order parameter become relevant on
energies scales of the order of the excitation gap $\omega \sim
\Delta$ (or, equivalently, for wavevectors $a|q_\nu| \sim
\Delta/v_\nu$) and qualitatively change the behavior of the gapless
modes at these energy scales (Appendix B). The description of the
latter via the non-linear sigma models in Eq.\ (\ref{Snls}) is
consequently restricted to small temperatures, small energies, and
large wavelengths.\cite{Tsve95}

As the gap decreases while the system approaches the phase transition
thermal fluctuations become sufficiently large to excite the amplitude
mode which in turn couples back to the angular Goldstone
modes. Amplitude fluctuations at energy scales of the order of or
larger than the gap can be treated outside the critical region by the
standard approach\cite{LV02} where Gaussian fluctuations are
integrated out. In Appendix B2 it is shown that the Gaussian
amplitude fluctuations can be included in the description of the
$\obOmega_{\mu}$ modes in Fourier space in form of a cutoff and a
fluctuation dependent prefactor: 
\begin{equation}\label{SOmegaq}
S_{\Omega}^{\rm amp} \approx 
\frac{\beta N}{4 \pi^2} \langle X^2 \rangle \!\!
 \sum_{\nu,\mu,n}^{0<\omega_n<\Delta} \!\!\!\! A_{\nu} \!\!\!\!
\int\limits_{0}^{\Delta/v_\nu}\!\!\!\!\! d^2 [a q]
    \left[\frac{\omega_n^2}{v_{\nu}^2}
 + [a q_\nu]^2 \right] \obOmega_{\mu}^2\,.
\end{equation} 
The number of in-plane Ru ions is $N=L^2/a^2$ and the prefactor
$\langle X^2 \rangle$ is the variance of the amplitude fluctuations
and parameterizes the effect of the amplitude fluctuations on the
action of the Goldstone modes. Following the analysis in 
Appendix B one can approximate $\beta \langle X^2 \rangle \approx
\chi^{(P)}_{\rbx{\rm BCS}}(0,0) = \frac{f_0}{\Delta}$ [Eq.\
(\ref{XX})]. Near the phase transition, where $\Delta \to 0$, the
prefactor diverges which is reminiscent of the diverging correlation
length.\cite{Schu95} For $T \to T_{\rm c}$ one has $S_{\Omega}^{\rm amp} 
\gg S_{\Omega}$ and the impact of the fluctuations of the internal
degrees of freedom of the order parameter is determined by Eq.\
(\ref{SOmegaq}). 

A dimensional analysis suggests that the integrals in Eq.\
(\ref{SOmegaq}) are proportional to $\Delta^5$. Using $A_\nu \approx
\frac{\ov_{\rm eff}^2}{\Delta f_0}$ [Eq.\ (\ref{Anu})], and $v_\nu
\approx \ov_{\rm eff}$ [Eq.\ (\ref{vnu})] the contribution to the action
from the amplitude-Goldstone mode coupling term can thus be estimated
to be   
\begin{equation}\label{SOmegasaddle} 
S^{\rm amp}_\Omega \approx  \frac{s_0^2\, \beta N}
{4\pi^3\, \ov_{\rm eff}^2}\ \Delta^3 + {\rm O}\left(\Delta^5\right) \,.
\end{equation}
The approximation is justified when the stiffness $\rho_s^{\rm amp}
\approx \beta \langle X^2 \rangle A_\nu \approx T \frac{\ov_{\rm
    eff}^2} {\Delta^2}$ is large enough, i.e., in the limit $\Delta\to
0$. In this approximation all channels $\mu$ yield the same 
contribution. The dimensionless prefactor $s_0^2 \sim 1$ is
phenomenological since the exact numerical prefactor is not known
within the approximations made. 

Equation (\ref{SOmegasaddle}) is an effective saddle point
contribution in which the fluctuations of the internal degrees
of freedom of the order parameter have effectively been integrated
out. The approach is only applicable outside the critical region $T
\le T_{\rm c} - 0.02$ K (Appendix B3).

The contribution from Eq.\ \ref{SOmegasaddle} is of third order in the
gap $\Delta^3$ near the phase transition. This an important
consequence of the two dimensionality of the problem. In one or three
dimensions the contribution is $\Delta^2$ or $\Delta^4$, respectively,
and consequently only renormalizes the usual parameters\cite{LL88} of
a Landau expansion.


\subsection{Free energy}\label{sectionFree} 

The general expression of the total free energy consists of the
contributions from the spin, spin-flavor, and charge channels and the
mean-field contribution, i.e., $F=\sum_\mu F_\mu + F_{\rm mf}$, with
\begin{eqnarray}\label{Fphi}
F_\mu = -T \ln\! \int\!\! d\varphi_\mu\,   
       {\rm e}^{-S_{\mu}} &\quad{\rm for}\quad& \mu\in\{{\rm sf},{\rm c}\},
\\  \label{Ftheta}
F_\mu = -T \ln\! \int\!\! d\vartheta_\mu\, 
       {\rm e}^{-S_{\mu}} &\quad{\rm for}\quad& \mu\in\{{\rm s}\},
\end{eqnarray}
where $d\varphi_{\rm c}=d\phi_\rho\,d\phi_{\rm f}$. The actions
$S_\mu$ are given by Eqs.\ (\ref{Seffsf}), (\ref{Seffs}), and
(\ref{Sc}) with $\langle P \rangle = 0$ in the normal phase. The
mean-field contribution is      
\begin{equation}\label{Fmf}
F_{\rm mf} = - V_{0} N \oP^2 -
T \ln\!\int\! \left[\prod\nolimits_{\bq,n}\!\! 
\frac{d\obOmega_{\rm f}} {(2\pi)} 
\frac{d\obOmega_{\rm s}} {(4\pi)} \right] {\rm e}^{-S_{\Omega}}.
\end{equation}
The second term accounts for the gapless fluctuations of the internal 
degrees of freedom of the order parameter (Sec.\ \ref{sectionfluct}
and Appendix).

The order parameter is obtained by minimizing the free energy via the
variation $(\delta F) / (\delta \oP) = 0$. At finite temperatures the
analysis of the expectation values of the sine-Gordon and related
models is quite involved.\cite{LM99} An estimate of the symmetry of
the full Eliashberg equation is discussed phenomenologically in Ref.\
\onlinecite{Wern02e}. Based on the result that interaction effects can
render the gap function rather homogeneous in agreement with the
experimental results\cite{Wern02c} (Sec.\ \ref{sectionsymm}) one can
derive a Landau theory for the free energy near the phase transition
in the ordered phase. 

To this end we omit at the moment the contribution from the
fluctuation action $S_{\Omega}$, assume an isotropic order
parameter,\cite{Wern02e,Wern02c} and consider the resulting
self-consistency equation which reads   
\begin{equation}\label{OP}
\oP =
\frac{2}{L^2 V_{0}} \sum\nolimits_\mu \frac{\delta F_\mu}{\delta \oP}
\,.
\end{equation}
To obtain an estimate for Eq.\ (\ref{OP}) we make use of the
equivalence of the actions Eq.\ (\ref{Seffsf}) and (\ref{Seffs}) to
those of spin chains with alternating interactions.\cite{ETD97} The
self-consistency Eq.\ (\ref{OP}) is thus equivalent to that studied in
the context of the spin-Peierls transition in CuGeO$_3$ studied in
Ref.\ \onlinecite{KRS99}. The results show that a one-dimensional
system that undergoes a phase transition to long range order---induced
by an effective three-dimensional coupling that can be treated
mean-field like---can be described near the phase transition by the
standard mean-field free energy functional\cite{LL88} 
\begin{eqnarray}\label{FLandau0}
\left[\sum\nolimits_\mu F_\mu + F_{\rm mf}\right]_{T\sim T_{\rm c}} &=& 
 \sum\nolimits_\mu F_{0,\mu} 
\nonumber\\&&\hspace{-6ex}
+\ N V_{0} \left[\frac{T}{T_{\rm
                       c}} - 1\right] \oP^2 + N b\, \oP^4
\end{eqnarray}
and consequently 
\begin{equation}\label{FmuLandau}
\left[\sum\nolimits_\mu F_\mu \right]_{T\sim T_{\rm c}} = 
\sum\nolimits_\mu F_{0,\mu} + N V_{0} \frac{T}{T_{\rm
                       c}}  \oP^2 + N b\, \oP^4 .
\end{equation}
By analogy we adapt this description near the phase transition while
for $T\to 0$ the free energy saturates as\cite{Baxt82,LZ97,Zamo95}
\begin{equation}\label{Fmu0}
F_\mu(T=0) \sim \oP_{0}^{\frac{1}{1-(B_\mu^2)/(8\pi)}}\,.
\end{equation}
The dimerized spin chain is SU(2) invariant with $B^2_\mu =
2\pi$.\cite{Cole76,CF79,Hald82b} Since $B^2_\mu=2\pi$ is not a
singular point in the analysis\cite{LZ97} at $T=0$ similar behavior
for other values of $B_\mu$ can be expected at finite temperature. The
comparison of Eqs.\ (\ref{FmuLandau}) and (\ref{Fmu0}) yields an
effective finite temperature interaction parameter of 
\begin{equation}\label{BmuofT}
B^2_\mu(T\sim T_{\rm c}) \approx 4\pi \,,
\end{equation}
which is the value for free massive fermions.\cite{GNT98}
Near $T_{\rm c}$ Eq.\ (\ref{DeltaGap}) then yields a gap of\cite{LZ97}   
\begin{equation}\label{DeltaofP}
\Delta \approx 8\ V_{0}\ \oP  \,.
\end{equation}

As discussed in Sec.\ \ref{sectionSaddle} the the gapless fluctuation
contribution at the phase transition can be described via the
effective saddle point term Eq.\ (\ref{SOmegasaddle}), i.e., 
$S_{\Omega} \to S^{\rm amp}_\Omega$ in Eq.\ (\ref{Fmf}). Together with
Eq.\ (\ref{FLandau0}) the Landau expansion of the free energy
$F_\Delta$ near the phase transition then is   
\begin{equation}\label{FLandau}
\frac{F_{\Delta}}{N} = \frac{\sum_\mu\!\! F_{0,\mu}}{N} 
     -V_{0}\,  t\, \oP^2\!
     + \frac{4 V_{0}^3 s_0^2}{\ov_{\rm eff}^2}\, 
                                \oP^3 \! +  b\, \oP^4 + \ldots\,,
\end{equation}
where $t=1-T/T_{\rm c} \ll 1$ is the reduced temperature. The approach
is valid outside the critical region, i.e., for $|t| \ge 0.01$ as
estimated in Appendix B3. Since the expansion is in $\oP\ge 0$
and {\em not} in the complex order parameter the phase transition
described by Eq.\ (\ref{FLandau}) is of third order in the sense of
Ehrenfest's definition.\cite{Ehre33} Gauge invariance is preserved
since $\oP = |e^{i\phi_{\rm G}}\langle P \rangle |$ for arbitrary
gauge fields $\phi_{\rm G}$.


\section{Application to S\lowercase{r}$_2$R\lowercase{u}O$_4$}
\label{sectionapplic} 


The inter-plane coupling in Sr$_2$RuO$_4$ leads to a superconducting
ground state driven by the $d_{zx}$ and $d_{yz}$ correlations. The
$d_{zx}$ and $d_{yz}$ correlations are described by the model derived
in Ref. \onlinecite{Wern02a} and applied in Sec.\
\ref{sectionmicro}. The pair order parameter is the 
expectation value of orbital-singlet, spin-triplet Cooper pairs. The
gapless fluctuations of the multiple order parameter components are
dominantly two dimensional. In this section the physical implications
of the model are compared with the properties of Sr$_2$RuO$_4$
determined through experiments.


\subsection{Order parameter: $\mu$SR and excess tunneling
current}\label{sectionOP}

\begin{figure}[bt]
   \epsfxsize=0.48\textwidth
   \centerline{\epsffile{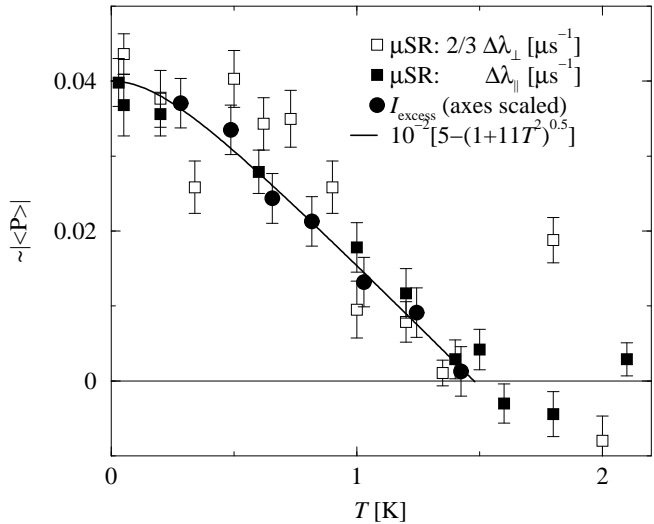}}
   \centerline{\parbox{\textwidth}{\caption{\label{OPcomp}
   \sl Order parameter as a function of temperature. Circles are the
   excess current form Ref. \protect\onlinecite{LGL+00} (axes scaled),
   squares are $\mu$SR relaxation rates from Ref.\
   \protect\onlinecite{LFK+98} with subtracted normal phase relaxation
   rate. Full squares are for Muon spins polarized in the $x$-$y$
   plane, open squares are scaled with 2/3 for Muon spins polarized
   perpendicular to the $x$-$y$ plane. The phenomenological fit
   $10^{-2} (5-\sqrt{1+11T^2})$ [Eq.\ (\ref{DeltaofT})] is consistent
   with the linear temperature dependence of the order parameter
   predicted in Eq.\ (\protect\ref{OPLandaut}).}}}     
   \end{figure}

Minimizing the Landau expansion of the free energy Eq.\
(\ref{FLandau}) via $(\partial F_\Delta) / (\partial \oP) = 0$ 
gives     
\begin{equation}\label{OPLandaut}
\oP\big|_{t\ll 1}\! = \frac{3 V_{0}^3 s_0^2}
                         {2\, b\, \ov_{\rm eff}^2} 
\!\left[\sqrt{1 + 
\frac{32\, b\, \ov_{\rm eff}^4\ t}{9\, V_{0}^5 s_0^4}} 
                                                     - 1\right] 
\approx 
\frac{\ov_{\rm eff}^2}{6\, V_{0}^2 s_0^2}\, t\,.
\end{equation}
The slope $\frac{\ov_{\rm eff}^2}{6\, V_{0}^2 s_0^2} \approx
0.1$ has been determined from the analysis of the specific heat data
in Sec.\ \ref{sectionC}. For $T\to 0$ or $t\to 1$ Nernst's theorem
requires the order parameter to saturate. A corresponding fit function
to the experimental data as shown in Fig.\ \ref{OPcomp} yields using
Eq.\ (\ref{DeltaofP})  
\begin{equation}\label{DeltaofT}
\Delta(T) = \left(5-\sqrt{1+11T^2/{\rm K}^2}\right)\ {\rm K}
\end{equation}
and is consistent with Eq.\ (\ref{OPLandaut}) for $T_{\rm c}=1.48$
K. Note that the determination of $T_{\rm c}$ from experiments for an
order parameter linear in $t$ is slightly different than in the case
without gapless fluctuations where $\Delta\sim\sqrt{t}$, see Sec.\
\ref{sectionC}. 

The squares in Fig.\ \ref{OPcomp} show the increase of the $\mu$SR
relaxation rates $\Delta\lambda_{\bot}$ (open symbols, Muon spin
polarized perpendicular to the $x$-$y$ plane) and $\Delta\lambda_{||}$
(full symbols, Muon spin polarized parallel to the $x$-$y$ plane)
below the superconducting phase transition from Ref.\
\onlinecite{LFK+98}. The increase of the $\mu$SR relaxation rate is a
consequence of the gapless magnetic fluctuations induced by the
spin-triplet Cooper pairs in the superconducting phase and described by
$\obOmega_{\rm s}$. The magnetic excitations can scatter dynamically
off the Muon spins. The resulting coupling to the Muon spin must be
described in the dynamical limit as\cite{Slic90} 
\begin{equation}
\lambda_s=B_0^2\ \tau^{-1}_{\rm s}\,.
\end{equation}
$B_0$ is the field at the Muon site, $\tau^{-1}_{\rm s}$ is the 
Cooper pair spin scattering rate. From the absence of any change in
the static susceptibility\cite{IMK+98,DHM+00} can be concluded that
$B_0$ is the same in both the normal phase and the superconducting
phase. The absence of the $\obOmega_{\rm s}$ fluctuations in the
normal phase yield $\lambda_s \gg \lambda_n$, where $\lambda_n$ is the
Muon relaxation rate induced by the ``normal'' electrons. Since the
contribution from the Cooper pairs is $\sim \lambda_s \oP/\oP_0$ while
that from the ``normal'' electrons is $\sim \lambda_n (1 - \oP/\oP_0)$
one has 
\begin{equation}
\Delta\lambda = (\lambda_s - \lambda_n)\ \oP / \oP_0 \,.
\end{equation}
The results from $\mu$SR are consistent with the linear temperature
dependence of the order parameter (Fig.\ \ref{OPcomp}).

In the presence of an external magnetic field the magnetic components
of the order parameter are Zeeman split as discussed in Sec.\
\ref{sectionfields}. The spin scattering rate $\tau^{-1}_s$ and thus
$\lambda_s$ are exponentially suppressed. This is consistent with the
experimental observation.\cite{LFK+98} 

As mentioned in Sec.\ \ref{sectionfluct} the spin-flavor coupling
terms in Eq.\ (\ref{Hintz}) could in principle lead to a gap in the
magnetic excitation spectrum. On the other hand such a coupling would
inevitably add a magnetic component to the flavor degrees of
freedom. Since the SO(2) flavor fluctuations remain
gapless\cite{Volo92} the Muon spin would then couple dynamically to 
the magnetic moment of the flavor mode leaving the physical picture of
the origin of the increase of the $\mu$SR relaxation rate unaltered.

From $\Delta\lambda_{||}\approx 2/3 \Delta\lambda_{\bot}$ it can be
concluded that the spin-one Cooper pairs are in a slight easy plane  
configuration, i.e., $B_{\bot}\approx 0.8 B_{||}$. Note that
$\Delta\lambda_{||}\sim B_{\bot}^2$ and vice versa.\cite{Slic90} From
the critical field analysis in Ref.\ \onlinecite{Wern02c} the in-plane
moment can be determined more precisely to be enhanced by a factor
1.23 with respect to the out-of-plane moment.

The excess current across a normal-superconducting point contact is 
proportional to the superconducting gap at least in $s$-wave
superconductors.\cite{BTK82} As discussed in Sec.\ \ref{sectionsuper}
and Ref.\ \onlinecite{Wern02e} the order parameter in Sr$_2$RuO$_4$
has extended $s$-wave character and in Sec.\ \ref{sectionsymm} the
superconducting gap in Sr$_2$RuO$_4$ is determined to be 93\%
isotropic. The full circles in Fig.\ \ref{OPcomp} are excess current
data from Ref.\ \onlinecite{LGL+00} and are in striking agreement with
the linear temperature dependence of the order parameter. The data are
scaled to match $T_{\rm c}$ and the slope.

Equation (\ref{DeltaofT}) gives an energy gap of $\Delta|_{T\to
0}\approx 0.33$ meV. The results from the differential resistance
versus bias voltage measurements across point contact junctions depend
on the model used for their analysis. A $p$-wave analysis without
gapless excitations yields $\Delta|_{T\to 0}\approx 1.1$ meV, a
$s$-wave analysis $\Delta|_{T\to 0}\approx 0.25$  meV.\cite{LGL+00}
Both results are close to the value extracted here. 

The differential resistance data  have been argued to be inconsistent
with an isotropic $s$-wave superconductor in the absence of gapless
excitations.\cite{LGL+00,MNJ+01} The theoretical approach in Ref.\
\onlinecite{SKY02} supports this conclusion but ignores contributions
from the $d_{zx}$ and $d_{yz}$ orbitals. An analysis of the Andreev
reflection spectroscopy data in the presence of the herein discussed
order parameter including massless fluctuations is desirable. 

Within the $p$-wave approach proposed in the literature
\cite{RS95,SAF+99} the order parameter $\bd(\bk)=\hat{\bz}(k_x\pm i
k_y)$ breaks time reversal symmetry and thus can also explain the
change in the Muon relaxation rate\cite{LFK+98,MRS01} at the
superconducting phase transition. The resulting temperature dependence
of the order parameter enters via the local field and is consequently
$\sim \sqrt{t}$ consistent with the differential resistance
analysis\cite{LGL+00} but not with the measured linear temperature
dependence of the excess current. On the other hand, the excess
current might be dominated by surface effects not considered here. A
more obvious possible shortfall of the $p$-wave approach are the
neutron scattering results\cite{BSB+02} that do not show any signature
of a gap down to energy transfers of $\omega \ge 1$ meV suggesting
that $2\Delta < 1$ meV which is inconsistent with the gap
value\cite{LGL+00} of $\Delta|_{T\to 0}\approx 1.1$ meV discussed
above.


\subsection{Specific heat}\label{sectionC}

\begin{figure}[bt]
   \epsfxsize=0.48\textwidth
   \centerline{\epsffile{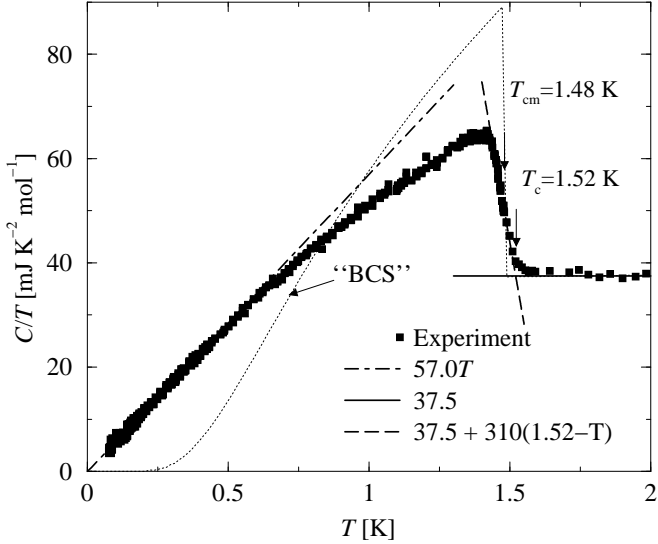}}
   \centerline{\parbox{\textwidth}{\caption{\label{CP148}
   \sl Specific heat over temperature as a function of
   temperature. Symbols are from Ref.\
   \protect\onlinecite{NMM00}. $T_{\rm cm}=1.48$ K is the ``mid
   transition'' value. The actual transition is at $T_{\rm c}=1.52$
   K. In the normal phase the specific heat is linear (full line)
   consistent with Fermi or Luttinger liquid
   behavior.\protect\cite{Wern02a} For $T$ just below $T_{\rm c}$ a
   linear increase with $t=1-T/T_{\rm c}$ is observed consistent with
   an order parameter linear in $t$ [Eq.\ (\ref{csTc})]. For $T<0.3$ K 
   the specific heat is quadratic in temperature [dot-dashed line,
   Eq.\ (\ref{cslowT})]) consistent with two-dimensional, gapless
   order parameter fluctuations. Dotted line: sketch of the BCS
   contribution.}}}    
   \end{figure}

The normal phase thermodynamic properties have been described
satisfactorily in the framework of the model outlined in Ref.\
\onlinecite{Wern02a}. The specific heat $C_n^{\rm tot}\approx 37.5\
\frac{\rm mJ}{\rm K^2 mol}\ T$ is linear in temperature for $T_{\rm c}
< T < 30$ K as is shown by the solid line in Fig.\ \ref{CP148}.

In the superconducting phase near the phase transition, i.e.,
$t=1-T/T_{\rm c}\ll 1$, the free energy is given by $F_\Delta$ in Eq.\
(\ref{FLandau}) while the temperature dependence of the
superconducting pair density $\oP$ is given by Eq.\
(\ref{OPLandaut}). The resulting specific heat $C_s=-T\partial^2
F_\Delta/(\partial T^2)|_{t>0}$ is  
\begin{equation}\label{csTc}
\frac{C_s|_{t\ll 1}}{T_{\rm c}}= \frac{C_n^{\rm tot}}{T_{\rm c}} + 2
\left[\frac{\ov_{\rm eff}^2}{6\ V_{0}^2 s_0^2}\right]^2
\frac{V_{0}}{T_{\rm c}^2}\ t - {\rm O}(t^2)\,.
\end{equation}
For $V_{0}=6$ K (see Sec.\ \ref{sectionnormal}) and $T_{\rm c}=1.52$ K
according to the fit in Fig.\ \ref{CP148} (dashed line) one has
$\frac{\ov_{\rm eff}^2}{6\, V_{0}^2 s_0^2} \approx 0.1$
which is consistent with the measured superconducting energy gap as
discussed in Sec.\ \ref{sectionOP}. One finds $\ov_{\rm
eff}^2/s_0^2\approx 22$ K$^2$. Note that the ``mid transition''
temperature $T_{\rm cm} = 1.48$ K is slightly lower that the critical
temperature $T_{\rm c} = 1.52$ K determined here.  

For slightly lower temperatures, i.e., $t>0.05$, the negative, higher
order terms in Eq.\ (\ref{csTc}) quickly become dominant. For
temperatures $0.7>t>0.05$ the experimental specific heat is given 
by a superposition of the exponential term from the gapped electronic
charge, spin, and spin flavor channels $F_\mu$ [Eqs.\
(\ref{Fphi}) and (\ref{Ftheta})] and from the non-linear sigma model
[Eq.\ (\ref{SOmegaq})]. 

The appropriate sine-Gordon actions for the electronic channels are
given by Eqs.\ (\ref{Seffsf}), (\ref{Seffs}), and (\ref{Sc}). With
finite $M_\mu\neq 0$ they yield a BCS-like, exponential specific heat
contribution as sketched be the dotted line in Figure
\ref{CP148}. The fluctuation contribution from the non-linear sigma
model Eq.\ (\ref{SOmegaq}) to the specific heat in Eq.\ (\ref{csTc})
is negative $\sim -t$ for $t\to 0$. Consequently the gapless
fluctuations account for the jump in the specific heat at $T_{\rm c}$
that is lower than anticipated by a gapped mean-field
calculation.\cite{NMM00} The physical interpretation is that a
significant contribution to the entropy is contained in the gapless
fluctuations of the internal degrees of freedom of the order
parameter. 

For small temperatures $t > 0.8$ the contribution from the electronic
channels is negligible\cite{NMM00} and the quadratic temperature
dependence of the non-linear sigma model Eq.\ (\ref{Snls}) dominates
the specific heat. 
\begin{equation}\label{cslowT}
C_s|_{t \to 1}= 3\ \frac{3\,\zeta(2)}{\pi}\ \frac{T^2}{(v_\nu)^2}
\approx 3.4\  \frac{T^2}{(\ov_{\rm eff})^2}
\end{equation}
Here $\zeta(n)$ is Riemann's zeta function and the prefactor 3 is the
multiplicity of the channels $\mu={\rm s}, {\rm f}$ that
contribute.\cite{Magnonquote} From the fit in Fig.~\ref{CP148}
(dash-dotted line) one extracts $\ov_{\rm eff}\approx 22$ K. The
resulting value of $s_0 \approx 4.7$ is only an order of magnitude
estimate since the accurate prefactor in Eq.\ (\ref{SOmegasaddle}) is
not known. If one assumes the magnetic channel to be gapped the
numbers are $\ov_{\rm eff}\approx 38$ K and $s_0 \approx 2.7$.

The value of $2\,\ov_{\rm eff}\approx 44 - 76$ K must be compared with
the quasi one-dimensional magnetic excitation velocity determined in
Ref.\ \onlinecite{Wern02a} as $v_{\rm eff} \sim 10^2$ K albeit outside
the Fermi liquid regime. The discrepancy must be attributed to effects
of the two-dimensional coupling and the pair correlations in the
superconducting state. The fact that the numbers are of the same order
of magnitude is a non-trivial consistency check of the theory.

The specific heat has been reproduced fairly well within the $p$-wave
approach by different groups\cite{NMM00,ZR01,ALGW01} by modeling the
gapless excitations via introduced line nodes or a two-gap
scenario,\cite{KS02} where one of the gaps is very small. Since a
number of experimental probes require a rather homogeneous in-plane
gap\cite{LGL+00,TSN+01,ITY+01,YAM+01} only horizontal line nodes are
possible. The latter require the fine tuning of a number of
parameters.\cite{ZR01,KS02,HS01,ALGW01} The finite slope near $T_{\rm
  c}$ has not been addressed yet. It would be interesting to repeat
those approaches using the pairing potential in Eq.\ (\ref{Vq})
including the gapless fluctuations of the internal degrees of freedom
of the order parameter.

\subsection{Stiffness and effective saddle-point regime}

The stiffness near the phase transition, where the amplitude-Goldstone
mode coupling is relevant, has been determined in Sec.\
\ref{sectionSaddle} as $\rho_s^{\rm amp} \approx \beta \langle X^2
\rangle A_\nu \approx T \frac{\ov_{\rm eff}^2} {\Delta^2}$. At low
temperatures it has be determined within the non-linear sigma model Eq.\ 
(\ref{Snls}) as $\rho_s^{0} \sim \oP^2 A_\nu |_{T\to 0}$. Since at low
temperatures the exact relation between $\oP$ and the gap is not known
[c.f.\ Eq.\ (\ref{DeltaGap})] a precise numerical determination of
$\rho_s^{0}$ is not possible. For sufficiently strong coupling the
reasonable assumption is that the stiffness is of the order of the
typical energy scale of the system, i.e., $\rho_s^{0} \approx \ov_{\rm
  eff}$ (c.f.\ Ref.\ \onlinecite{Schu95}). 

\begin{figure}[bt]
   \epsfxsize=0.48\textwidth
   \centerline{\epsffile{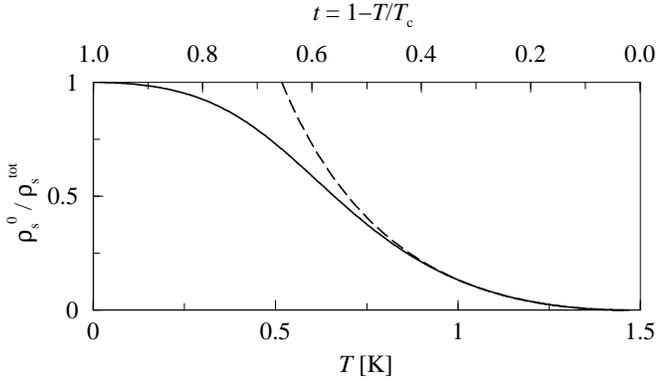}}
   \centerline{\parbox{\textwidth}{\caption{\label{stiffps}
   \sl Normalized Inverse stiffness $\rho_s^{0} / \rho_s^{\rm tot}(T)$
   (full line) from Eq.\ (\ref{rhotot}) for $\rho_s^{0} \approx
   \ov_{\rm eff}$. The broken line shows the asymptotic part $\ov_{\rm
   eff} / \rho_s^{\rm amp}$.}}} 
   \end{figure}

To obtain a qualitative picture of the temperature dependence of the
stiffness the approximate interpolation formula 
\begin{equation}\label{rhotot}
\rho_s^{\rm tot}(T) \approx 
  \sqrt{(\rho_s^{0})^2 + (\rho_s^{\rm amp})^2} \approx 
  \ov_{\rm eff}
\sqrt{ 1 + \frac{T^2\,\ov_{\rm eff}^2}{\Delta^4}}
\end{equation}
can be applied. The full line in Fig.\ \ref{stiffps} shows the
normalized inverse stiffness from Eq.\ (\ref{rhotot}), the broken line
is the asymptotic part $\ov_{\rm eff} / \rho_s^{\rm amp}$. The
temperature dependence of $\Delta(T)$ has been modeled via Eq.\
(\ref{DeltaofT}).

At intermediate temperatures $0.8 > t > 0.05$ the stiffness increases
as $\sim t^{-2}$ and the integrals in the non-linear sigma models need
to be cut off as in Eq.\ (\ref{SOmegaq}). For $t < 0.05$ one has
$\ov_{\rm eff} / \rho_s^{\rm amp} \le 10^{-3}$ which is sufficiently
small to justify the effective saddle point approach to the fluctuation
action Eq.\ (\ref{SOmegasaddle}) and the resulting Landau expansion of
the free energy Eq.\ (\ref{FLandau}). 

For $t > 0.05$ the effective saddle point approximation Eq.\
(\ref{SOmegasaddle}) is not applicable any more. The third order term
in Eq.\ (\ref{FLandau}) then is not present in that form anymore which
enhances the importance of the higher order terms in Eq.\ (\ref{csTc})
and contributes to the rather sharp maximum of the specific heat. Note
that even in BCS theory the specific heat is much more sensitive to
the higher order terms that stem from the opening of the single
particle gap than the temperature dependence of the order parameter
itself.


\subsection{Knight shift, static susceptibility, NQR, and universal
  thermal transport}\label{sectionmag}

The explanation of the observed absence of a change in the Knight
shift\cite{IMK+98} and the magnetic susceptibility\cite{DHM+00} in
the superconducting phase is trivial within the present model. All
order parameter components carry spin one and thus no electronic
magnetic moments are lost in the superconducting phase.

The relaxation times determined by NQR are consistent with gapless,
two-dimensional fluctuations in the superconducting
state.\cite{IMK+00} The qualitative agreement with the model presented
here is obvious. 

The thermal transport also is consistent with the presence of gapless
fluctuations below $T_{\rm c}$.\cite{STK+02} The observed universal
linear temperature dependence for $T\to 0$ is determined by the long
wavelength excitations of the internal degees of freedom of the order
parameter and is consequently determined by the properties of
collective state much rather then by the number of impurities in the
system.


\subsection{Order parameter symmetry}\label{sectionsymm}

Equation (\ref{Omegaf}) reveals the spatial order parameter anisotropy 
\begin{equation}\label{Pprop}
\langle \bP \rangle \propto \langle\bOmega_{\rm f}\rangle = 
\left( \langle\sin (\sqrt{\pi} \phi_{\rm f} - 
                                 \sqrt{2}\, \oy) \rangle \atop 
       \langle\sin (\sqrt{\pi} \phi_{\rm f} - 
                                 \sqrt{2}\, \ox) \rangle \right)
\end{equation}
As discussed in Sec.\ \ref{sectionfields} the pair correlations for
$\langle \Omega_{{\rm f},x} \rangle$ are larger along $\ox$ or [110]
and for $\langle \Omega_{{\rm f},y} \rangle$ are larger along $\oy$ or 
[$\overline{1}$10]. The simplest parameterization is 
\begin{equation}\label{fparameter}
\langle\bOmega_{\rm f}\rangle
= \frac{\langle \Omega_{{\rm f},0}\rangle}{\sqrt{2}} 
        \left(1  \atop 1 \right) +
\sin 2\eta\ \frac{\langle \Omega_{{\rm f},1} \rangle}{\sqrt{2}}
 \left(1 \atop -1 \right)\,,
\end{equation}
where the angle $\eta$ is measured with respect to the [100] axis.
Each component has a two-fold symmetry while the resulting spatial
symmetry of the total order parameter reflects the four-fold axis of
the underlying lattice. 
\begin{equation}\label{fsymmetry}
\left|\langle\bOmega_{\rm f}\rangle\right|
=\sqrt{
\langle \Omega_{{\rm f},0} \rangle^2 +  
            \langle \Omega_{{\rm f},1}\rangle^2\ \sin^2 2\eta}
\end{equation}
Oriented magnetic fields and temperature gradients break the SO(2)
symmetry of $\bOmega_{\rm f}$ (Sec.\ \ref{sectionfields}) and reveal
the two-fold symmetry of the components. Thermal conductivity 
measurements\cite{ITY+01} suggest that $\langle \Omega_{{\rm
f},1}\rangle / \langle \Omega_{{\rm f},0} \rangle \approx 0.02$. A
more careful analysis of the upper critical fields\cite{YAM+01} in
Ref.\ \onlinecite{Wern02c}  yields $\langle \Omega_{{\rm f},1}\rangle 
/ \langle \Omega_{{\rm f},0} \rangle = 0.034$ which leads to a spatial 
anisotropy of the order parameter components of 7\%. The the four-fold
symmetry of the total order parameter is then  
\begin{equation}\label{fsymnum}
\left|\langle\bOmega_{\rm f}\rangle\right|
\approx \langle \Omega_{{\rm f},0} \rangle \left( 1+
      \frac{\langle \Omega_{{\rm f},1}\rangle^2}
           {\langle \Omega_{{\rm f},0} \rangle^2}\ \sin^2 2\eta \right)\,,
\end{equation}
which is also consistent with the experimental
observations.\cite{ITY+01} 

The experimentally implied existence of two order parameter components
with a slight spatial anisotropy\cite{Agte01} thus follows quite
naturally out of the degenerate saddle points in the flavor channel
discussed in Sec.\ \ref{sectionOrder}.


\section{Relation to the \lowercase{$p$}-wave approach}
\label{sectionDisc}

A large number of experiments in superconducting samples of
Sr$_2$RuO$_4$ have been interpreted in terms if the time reversal
symmetry breaking $p$-wave state with symmetry $E_u$ and order
parameter $\bd(\bk)= \hat{\bz}(k_x \pm i k_y)$.\cite{SAF+99,MRS01} As 
has been discussed throughout the manuscript most experimental results
find alternative interpretations within the present approach. Examples
are 

(i) the observed density of states below the superconducting
gap\cite{NMF+98,NMM99,NMM00,STK+02,IMK+00} that has persistently been
interpreted as an indication for line-nodes in the gap function while
it may well originate from any kind of gapless, two-dimensional
excitations with linear dispersion (Secs.\ \ref{sectionfluct},
\ref{sectionC}, and \ref{sectionmag});

(ii) the $c$-axis pressure dependence of $T_{\rm c}$ that
was interpreted in Ref.\ \onlinecite{OSM+02} as a consequence of a
shift of the $\gamma$ band closer to the van Hove singularities near
the $M$ point while in Sec.\ \ref{sectionmf} it is viewed as the
simple enhancement of the inter-plane coupling;

(iii) the strong in-plane coupling of strain to the superconductivity
which was interpreted as a possible indication of an oder parameter of
the form $\bd(\bk)= \hat{\bz}(k_x \pm i k_y)$ while the anisotropy of
the components of $\bOmega_{\rm f}$ can account for the same effect for
symmetry reasons (Sec.\ \ref{sectionfluct});

(iv) the change in the $\mu$SR relaxation rate across the
superconducting phase transition that was interpreted as an indication
for a time-reversal symmetry breaking pairing
state\cite{LFK+98,SAF+99} while it may as well stem from the
scattering of gapless magnetic excitations off the Muon spin (Sec.\
\ref{sectionOP});

(v) the sensitivity of the superconductivity to non-magnetic impurities
which has been interpreted as an indication for higher angular momentum
pairing states\cite{NMM99,MHT+98,MMM99} while it follows directly out
of the symmetry of the orbital-singlet pairing state as discussed in
Ref.\ \onlinecite{Wern02e}.

These similarities between the approaches appear not very surprising
when considering that the spin-flavor phase space discussed herein
includes a time-reversal pairing state similar to the $p$-wave $E_u$
state. The flavor expectation values $\langle \bOmega_{\rm f} \rangle$
in Eq.\ (\ref{Omegaf}) possibly contain contributions that are odd in
real space, i.e., $\langle \sin \sqrt{2}\oy\, \cos \sqrt{\pi}
\phi_{\rm f} \rangle \neq 0$ and $\langle \sin \sqrt{2}\ox\, \cos
\sqrt{\pi} \phi_{\rm f} \rangle \neq 0$, and consequently a
representation   
\begin{equation}\label{Omegap}
\langle\bOmega_p\rangle  \propto \left(
\begin{array}{c}
\langle \sin \sqrt{2}\oy\, \cos \sqrt{\pi} \phi_{\rm f} \rangle
+ i \langle \sin \sqrt{2}\ox\, \cos \sqrt{\pi} \phi_{\rm f} \rangle \\
\langle \sin \sqrt{2}\oy\, \cos \sqrt{\pi} \phi_{\rm f} \rangle
- i \langle \sin \sqrt{2}\ox\, \cos \sqrt{\pi} \phi_{\rm f} \rangle
\end{array}\right)
\end{equation}
can be constructed with $\bOmega_p \in \{\bOmega_{\rm f}\otimes
\bOmega_{\rm s}\}$. The magnetic configuration of $\bOmega_p$ depends
on the details of the spin-flavor coupling terms and an easy-plane
configuration as discussed in Sec.\ \ref{sectionOP} and in Ref.\
\onlinecite{Wern02c} is quite conceivable. The present approach with
mixed-orbital singlets can thus be generalized to include the
$p$-wave\cite{Sigr00,Agte01} $E_u$ picture.

The conceptual advantage of the present approach is its derivation
from a microscopic model. It allows to study interaction effects like
the presence of two degenerate saddle points in detail---at least on a
qualitative basis. Phenomenologically the important difference is that
here Hund's rule coupling is emphasized over effects from
spin-orbit coupling as discussed in Sec.\ \ref{sectionnormal} which
allows for the degeneracy in the internal degrees of freedom (Sec.\
\ref{sectionfluct}). This leads to a coherent description of a
multitude of observations in Sr$_2$RuO$_4$ including the experiments
discussed herein as well as the normal phase magnetism (Ref.\
\onlinecite{Wern02a}), the interplay of the lattice symmetry,
impurities and interaction effects (Ref.\ \onlinecite{Wern02e}), and
the magnetic field dependence (Ref.\ \onlinecite{Wern02c}).  

Resulting specific advantages over the $p$-wave $E_u$ description are 

(i) no need to construct horizontal line nodes or two-gap states which
require parameter fine tuning and which have not been confirmed
experimentally; 

(ii) no inconsistency between the gap value extracted from Andreev
reflection spectroscopy with respect to neutron scattering results
(Sec.\ \ref{sectionOP});

(iii) the correct two-component order parameter symmetry as determined
via thermal transport and upper critical field measurements (Sec.\
\ref{sectionsymm});

(iv) the description of the temperature slope of the specific heat
near $T_{\rm c}$ (Sec.\ \ref{sectionC}) as well as the magnetic field
dependence of the specific heat near $H_{\rm c}$ (Ref.\
\onlinecite{Wern02c}).  

To obtain a more conclusive discrimination of the state promoted in
this series of papers and the $p$-wave $E_u$ state we propose to study
a detailed Landau-Ginzburg functional\cite{Prot02} that contains in
its parameter space the degenerate state described herein as well as
the special case\cite{Sigr00,Agte01} given in Eq.\
(\ref{Omegap}). Experimentally, the presence of the $\obOmega_{\rm f}$
mode might be detectable via microwaves.


\section{Conclusions}

From the approach discussed in the present paper the essential
physical properties of the superconducting state in Sr$_2$RuO$_4$ can
be summarized as follows.

The body centered tetragonal structure gives rise to inter-plane pair
correlations in the $d_{\rm yz}$ and $d_{\rm  zx}$ orbitals enhanced
by umklapp scattering processes. The superconducting order parameter
is found to be a mixed-orbital-singlet spin-one-triplet. The bosonized
description of the in-plane electron correlations is consistent with
the order parameter and the slight easy plane configuration of the
magnetic moments. The model reveals that the oder parameter has two
slightly anisotropic spatial components. 

The absence of a change in the Knight shift and the magnetic
susceptibility in the superconducting phase follows trivially from the
spin-one Cooper pairs. The experimental thermal conductivity and upper
critical fields are consistent with a spatial anisotropy of the order
parameter components of 7\%. 

The different components of the order parameter in the spin and flavor
sector give rise to two-dimensional gapless fluctuations. They are
modeled by a 2+1 dimensional non-linear sigma model. Near the phase
transition the two-dimensional gapless fluctuations account for the
linear dependence of the pair density and the specific heat on the
reduced temperature. At low temperatures they yield the quadratic
temperature dependence of the specific heat, the cubed temperature
dependence of the NQR relaxation time, and the universal linear
temperature dependence of the thermal transport. 

The finite pair density of the electrons in the $d_{xy}$ orbitals is
induced by the inter-band proximity effect. There is only one phase
transition in the absence of magnetic fields and one single particle
gap.


\section*{Acknowledgments}

I am grateful to V.\ J.\ Emery for initiating this project. I thank
S.\ Carr, D.\ F.\ Agterberg, A.\ M.\ Tsvelik, M.\ Sigrist, B.\ O.\
Wells, J. Kroha, J. Brinckmann, M.\ Esch\-rig, K.\ Kikoin, H.-H.\
Klauss, M.\ Weinert, A.\ Kl\"umper, and H.\ Keiter for instructive
and stimulating discussions. I thank A.\ M.\ Tsvelik and G.\ E.\
Volovik for valuable comments on the spectra in triplet
superconductors and $^3$He-$A$. The work was supported by DOE contract
number DE-AC02-98CH10886 and the Center for Functional
Nano\-struc\-tures at the University of Karlsruhe.


\section*{Appendix: Derivations for the order parameter
  fluctuations}\label{sectionDerive} 
  
The expressions that are used in the manuscript to describe the
fluctuations of the internal degrees of the order parameter can be
derived within standard perturbative
approaches.\cite{Schu95,GNT98,Frad91,Schu90b,WTL91,LV02} Within the 
vectorial representations introduced in Sec.\ \ref{sectionOrder} the
pair operators in the 
relevant orbital-singlet spin-triplet channel are given as
\begin{equation}\label{POmega}
{\bP_{t}^{s}} = 
 \frac{2}{\pi}\, 
     \sin\sqrt{\pi}\phi_{\rm sf}\        
     \bOmega_{\rho}\otimes\bOmega_{\rm s}\otimes\bOmega_{\rm f}\,.
\end{equation}
Recall that the flavor degrees $\bOmega_{\rm f}$ of freedom stem from
the symmetry of the bosonized action which suggest a two-fold
degenerate saddle point. 

The superconducting state that is discussed in this paper involves the
breaking of various continuous symmetries. The broken gauge symmetry is
parameterized via the charge phase $\langle\bOmega_{\rho}\rangle$, the
SO(3) symmetry of the magnetic moment of the spin-triplet Cooper pairs
enters via $\langle\bOmega_{\rm s}\rangle$, and the broken SO(2)
symmetry of the flavor channel is given by $\langle\bOmega_{\rm
  f}\rangle$. The expectation value of the order parameter is given in
vectorial representation via 
\begin{equation}\label{POmegaexp}
\langle {\bP_{t}^{s}} \rangle = \oP\
     \obOmega_{\rho}\otimes\obOmega_{\rm s}\otimes\obOmega_{\rm f}\,,
\end{equation}
where $\oP$ is the order parameter amplitude as defined in Eq.\
(\ref{Paverage}) and $\obOmega_\mu(\br)$ are normalized according to
Eq.\ (\ref{Omegaaver}). The phase of $\langle\sin\sqrt{\pi}\phi_{\rm
  sf} \rangle$ can be absorbed in $\obOmega_{\rho}$ via a gauge 
transformation. 

The ground state is degenerate with respect to rotations of
$\obOmega_\mu$ giving rise to long wavelength gapless
excitations---the Goldstone modes. In order to describe the dynamics
of the system in the ordered phase in proves useful to introduce
pair operators that are related to those in Eq.\ \ref{POmega} by a
shift of the pair expectation value, i.e., $\tbP \equiv {\bP_{t}^{s}} -
\langle {\bP_{t}^{s}} \rangle$. The mean-field decoupling Eq.\
(\ref{mfdecoupling}) then can be rewritten as the exact relation   
\begin{equation}\label{fldecoupling}
{\bP_{t}^{s}}^\dagger{\bP_{t}^{s}}^{\phantom{\dagger}} = 
 {\bP_{t}^{s}}^\dagger \langle {\bP_{t}^{s}}^{\phantom{\dagger}} \rangle
+    {\bP_{t}^{s}}^{\phantom{\dagger}}
                        \langle {\bP_{t}^{s}}^{\phantom{\dagger}} \rangle^*
-    |\langle {\bP_{t}^{s}}^{\phantom{\dagger}} \rangle|^2 
+   \tbP^\dagger \tbP^{\phantom{\dagger}} .
\end{equation}
Using Eq.\ (\ref{fldecoupling}) the orbital-singlet spin-triplet part
of the pair Hamiltonian Eq.\ (\ref{HtperpFT}) decouples into two parts
$H_p = H_{\rm mf} + H_{\rm fl}$, where the mean-field part is given by
Eq.\ (\ref{Hmf}) or, in its bosonized version, by Eq.\
(\ref{Hmfeff}). The fluctuation part is  
\begin{equation}\label{Hfl}
H_{\rm fl} = 
  \sum_{\bq}\! \frac{V_{\bq}}{N}\
         \tbP^\dagger(\bq)\ \tbP^{\phantom{\dagger}}(\bq)\,.
\end{equation}

The partition function of the whole system can now be expressed via a
standard perturbative approach\cite{NO88,Schu90b,WTL91,Schu95,Volo92}
as  
\begin{equation}\label{Ztot}
Z = e^{-S_{\rm mf}}\ \int {\cal D}[\bX,\bX^*]\  
                      \exp\{-S_{\rm fl}[\bX,\bX^*]\}\,.
\end{equation}
The mean-field action is composed of Eqs.\ (\ref{Seffs}),
(\ref{Seffsf}), and (\ref{Sc}) as $S_{\rm mf} = S^{\rm eff}_{\rm s} + 
S^{\rm eff}_{\rm sf} + S_{\rm c}$ and accounts for the physics
contained in the saddle point including the mean-field temperature
dependence of the gap and the Goldstone modes as discussed in Appendix
A below. The thermodynamic averaging in the amplitude fluctuation part
$\sim S_{\rm fl}[\bX,\bX^*]$ is performed with respect to the
mean-field system $S_{\rm mf}$. The vectorial character of the pair
operators leads to Hubbard-Stratonovich
fields\cite{Schu90b,WTL91,Schu95} $\bX$ and $\bX^*$ which also have
vectorial character and can be parameterized as  
\begin{equation}\label{defX}
\bX = X\ \obOmega_{\rho}\otimes\obOmega_{\rm s}\otimes\obOmega_{\rm f}\,.
\end{equation}
The integral measure contains the averaging over the angular
components of the Hubbard-Stratonovich
fields\cite{Schu90b,WTL91,Schu95}
\begin{equation}\label{defD}
{\cal D}[\bX,\bX^*] = \prod_{\bq,\omega_n} 
           \sqrt{\beta |V_{\bq}|}\,
 \frac{d X_{\bq,n}}{\pi}\,
 \frac{d\obOmega_{\rho}}{2\pi}\,
 \frac{d\obOmega_{\rm s}}{4\pi}\,
 \frac{d\obOmega_{\rm f}}{2\pi}\,.
\end{equation}
The $\omega_n$ are bosonic Matsubara frequencies. The averaging over
the angular components corresponds to an averaging over the degenerate
ground states of the system.\cite{Schu90b,WTL91}


\subsection{Goldstone modes}\label{sectionGoldstone}

In the present approach the Goldstone modes that correspond to the
broken local spin rotational, flavor, and gauge invariance are
implicitly contained in the mean-field system defined by the actions
$S_{\rm mf} = S^{\rm eff}_{\rm s} + S^{\rm eff}_{\rm sf} + S_{\rm
c}$. This becomes obvious when considering that the mass terms in
Eqs.\ (\ref{Seffs}), (\ref{Seffsf}), and (\ref{Sc}) are decreased
when the operator $\bOmega_\mu$ is not parallel to the mean direction
$\obOmega_\mu$. A direct derivation of the non-linear sigma model in
Eq.\ (\ref{Snls}) that describes the dynamics of the Goldstone modes
within the present approach would require non-Abelian bosonization in
order to maintain the symmetry of the Hamiltonian
explicitly.\cite{Tsve95,GNT98} In the light of the well established
results in the literature the involved explicit derivation herein
appears obsolete.   

A simple qualitative argument that makes the underlying physics
transparent to the reader shall be discussed instead. The finite local
angle $\eta_\mu(\br,\tau) = \angle[\bOmega_\mu(\br,\tau),
\obOmega_\mu(\br,\tau)]$ increases the energy\cite{LZ97} of the system
since the mass term determining the gap is lowered. Consequently there
is a restoring force which can be assumed linear in
$\eta_\mu(\br,\tau)$ if the angle is sufficiently small. Small
fluctuations of $\eta_\mu(\br,\tau)$ are consequently expected to be
harmonic. Their spectrum is gapless since the restoring force vanishes
for $\eta_\mu(\br,\tau)=0$. The excitation velocity should be of the same
order\cite{Schu90b,WTL91,SWZ89} as the velocity of the gapped
amplitude mode (Appendix B). Finally, the local constraint of
$|\obOmega_\mu(\br,\tau)|^2 = 1$ distinguishes the angular
fluctuations from other harmonic systems and leads to the effective
description via the non-linear sigma models given by Eq.\ (\ref{Snls})
in Sec.\ \ref{sectionfluct}.   

At sufficiently small temperatures and energy scales $T,\omega \ll
\Delta$ the stiffness of the non-linear sigma model describing the
angular fluctuations has been determined\cite{Tsve95,GNT98} to scale
with the square of the order parameter amplitude $\sim \oP^2$, albeit
with corrections\cite{WTL91,Schu95} that depend on the specific
correlations of the system. As will be discussed in Appendix B2
the gap in the ordered phase plays the role of a cutoff.\cite{Tsve95}
This cutoff becomes small as the gap becomes small and has important
physical implications.\cite{CHN89}


\subsection{Gaussian amplitude fluctuations}\label{sectionGauss}

The integral term in Eq.\ (\ref{Ztot}) contains the fluctuations
$\tbP$ of the system involving changes in amplitude of the order
parameter. The amplitude fluctuation part is expanded up to second
order\cite{Schu90b,WTL91} in the complex fields $\bX$ and $\bX^*$. 
Considering that per definition $\langle \tbP \rangle = 0$ one has 
\begin{eqnarray}\label{Sflraw}
S_{\rm fl}[\bX,\bX^*] &=& \nonumber \\ &&\hspace{-9ex}
\beta\sum_{\bq,\omega_n} |V_{\bq}|
  \left[1 - 
        |V_{\bq}|\, \chi^{(P)}_{\rbx{2,\pm}}(\bq,i\omega_n)\right]
        |\bX(\bq,i\omega_n)|^2 .
\nonumber \\ &&
\end{eqnarray}
For the relevant umklapp scattering near $\bq\sim 0$ the potential has
been set $V_{\bq} \approx -|V_{\bq}|$. The pair correlation
function $\chi^{(P)}_{\rbx{2,\pm}}$ needs to be calculated with
respect to $S_{\rm mf}$. Since $S_{\rm mf}$ describes a gapped system
in the ordered phase $\chi^{(P)}_{\rbx{2,\pm}}(0,0)\big|_{T<T_{\rm c}}
< \chi^{(P)}_{\rbx{2,\pm}}(0,0)\big|_{T=T_{\rm c}}$ and the action in
Eq.\ (\ref{Sflraw}) is stable.

Limiting the description to the low energy long wavelength properties
one may expand   
\begin{eqnarray}\label{chiexpand}
|V_{\bq}|^2\, \chi^{(P)}_{\rbx{2,\pm}}(\bq,\omega_n) &\approx&
|V_{0}|^2\, \chi^{(P)}_{\rbx{2,\pm}}(0,0) 
\nonumber\\&&\hspace{-17ex}
             +\ A_x q_x^2 + A_y q_y^2 + 
  \left(\frac{A_x}{(a v_x)^2}+\frac{A_y}{(a v_y)^2}\right)(i\omega_n)^2,
\end{eqnarray}
where $A_\nu$ and $v_\nu$ are defined after analytical continuation
$i\omega_n \leftrightarrow \lim_{\epsilon \to 0} (\omega + i\epsilon)$
in Eqs.\ (\ref{stiffness}) and (\ref{vOmega}), respectively. As
discussed in Sec.\ \ref{sectionPCE} one can assume for the low
temperature and low energy regime the system to be isotropic in the
$x$-$y$ plane with $A_\nu = A_x = A_y$ and $v_\nu = v_x = v_y$.
 
Using Eq.\ (\ref{chiexpand}) and transforming Eq.\ (\ref{Sflraw}) to
Euclidean space representation the action becomes 
\begin{eqnarray}\label{SEuclid}
S_{\rm fl} &=&   \int_{-L}^L\! d^2r \int_0^\beta d\tau\ |V_{0}|
  \left[1 - 
        |V_{0}|\, \chi^{(P)}_{\rbx{2,\pm}}(0,0)\right]
        X^2 
\nonumber\\&&\hspace{-2ex}+\
\int_{-L}^L\! d^2r \int_0^\beta d\tau
 \sum_{\nu=x,y} A_{\nu} \Bigg[
\frac{\left|{\partial_\tau}\, \bX \right|^2} 
   {(a v_{\nu})^{2}} +
\left|{\partial_\nu}\, \bX\right|^2 \Bigg] .
\nonumber\\
\end{eqnarray}
The $\bq$ dependence of the term linear in $|V_{\bq}|$ has been
neglected. Note that in Euclidean space $|\bX|^2\equiv X^2$ because of
the local constraints $|\obOmega_{\mu}(\br,\tau)|^2 \equiv 1$. Another
important consequence of the local constraint is that $\bOmega_{\mu} 
\partial_\tau \bOmega_{\mu} = \bOmega_{\mu} \partial_\nu \bOmega_{\mu}
= 0$. Consequently angular and amplitude contributions to the
derivatives in Eq.\ (\ref{SEuclid}) decouple for $\nu=x,y,\tau$ as 
\begin{equation}\label{derivatives}
\left|{\partial_\nu}\, \bX \right|^2 = 
               \left({\partial_\nu}\, \bX \right)^2 + 
  X^2 \sum_{\mu = \rho,\rm s,f}
           \left({\partial_\nu}\, \obOmega_\mu \right)^2\,.
\end{equation}
Now the different modes around the saddle point can be discussed in
detail.


\subsubsection{Amplitude mode}

The part of Eq.\ (\ref{SEuclid}) that describes the amplitude
excitations is in Fourier space
\begin{eqnarray}\label{Samp}
S_{\rm amp}[X,X^*] &=&
\beta \sum_{\bq,\omega_n}
  \Bigg[|V_{0}|\left(1 - |V_{0}|\,
    \chi^{(P)}_{\rbx{2,\pm}}(0,0)\right)  
\nonumber\\&&\hspace{-2ex}
+\! \sum_{\nu=x,y}\! A_\nu\! \left((a q_\nu)^2  +
        \frac{(i\omega_n)^2}{v_\nu^2}\right)\Bigg]
        |X(\bq,i\omega_n)|^2 \nonumber\\&&
\end{eqnarray}
and describes Gaussian fluctuations around the static saddle
point.

In order for the action Eq.\ (\ref{Samp}) to be consistent with the
mean-field pair correlation function estimated by Eq.\ (\ref{chiBCS})
one must require that  
\begin{equation}\label{XX}
\beta\langle X(\bq,i\omega_n) X(-\bq,-i\omega_n) \rangle = 
\chi^{(P)}_{\rbx{\rm BCS}}(\bq,i\omega_n)\,.
\end{equation}
After analytical continuation $i\omega_n \leftrightarrow
\lim_{\epsilon \to 0} (\omega + i\epsilon)$ and comparing Eqs.\
(\ref{Samp}), (\ref{XX}), and $\chi^{(P)}_{\rbx{\rm BCS}}$ in Eq.\
(\ref{chiBCS}) one can identify 
\begin{eqnarray}
\label{Anu}
A_\nu &\approx& \frac{\ov_{\rm eff}^2}{\Delta f_0}\,, \\
\label{vnu}
v_\nu &\approx& \ov_{\rm eff}\,,
\end{eqnarray}
as well as 
\begin{eqnarray}\label{Gapdef}
|V_{0}|\left(1 - |V_{0}|\,
    \chi^{(P)}_{\rbx{2,\pm}}(0,0)\right) \approx
\frac{\Delta}{f_0}\,.
\end{eqnarray}
The $\bq$ dependence of $f(\bq) \approx f(0) \equiv f_0$ has been
neglected. The dispersion of the amplitude mode then can readily be
read off from Eq.\ (\ref{Samp}) as \begin{equation}\label{oPmode}
\omega_{\rm amp}=\ov_{\rm eff} \sqrt{a^2 \bq^2 - \Delta^2/\ov_{\rm eff}^2}
\end{equation}
and has the desired shape with an excitation gap $\Delta$ underlining
the self-consistency of the approach. 

The identical self-consistent results are obtained by directly
identifying  
\begin{equation}\label{chiidentify}
 |V_{\bq}|^{-1} \left[1 - |V_{\bq}|\,
      \chi^{(P)}_{\rbx{2,\pm}}(\bq,i\omega_n)\right]^{-1}  
    \approx  \chi^{(P)}_{\rbx{\rm BCS}}(\bq,i\omega_n)
\end{equation}
in the action Eq.\ (\ref{Sflraw}) which again shows the stability of the
latter since $\chi^{(P)}_{\rbx{\rm BCS}}(0,0) > 0$.


\subsubsection{Amplitude-Goldstone mode coupling}

Starting from the fluctuation action Eq.\ (\ref{SEuclid}), including
the identity Eq.\ (\ref{derivatives}), using the identifications
Eqs. (\ref{Anu}) and (\ref{vnu}), and transforming to Fourier space
the action 
\begin{eqnarray}\label{SOmega}
S_{\Omega}^{\rm amp}[\bOmega_\mu,\bOmega_\mu^*] &=&
\beta
\sum_{\bq,\omega_n}\sum_{\nu=x,y} A_\nu  
\sum_{\mu=\rho,\rm s,f} X^2(\bq,i\omega_n)
\nonumber\\&&\hspace{-3ex}\times\
 \left((aq_\nu)^2  +
        \frac{(i\omega_n)^2}{v_{\nu}^2}\right)
        |\bOmega_\mu (\bq,i\omega_n)|^2
\end{eqnarray}
is obtained for the angular modes. Except for the prefactors it is
equivalent to the non-linear sigma model Eq.\ (\ref{Snls}) used to
describe the low temperature, low energy excitations of the angular
Goldstone modes [Appendix A, Sec.\ \ref{sectionfluct}].

The action Eq.\ (\ref{SOmega}) is the lowest order amplitude-Goldstone
mode coupling term. This becomes evident when replacing the amplitude
of the pair fluctuation field $X$ by its expectation value, i.e., $X\to
\langle X \rangle = 0$: in the absence of amplitude fluctuations Eq.\
(\ref{SOmega}) does not contribute. This is reminiscent of the
decoupling of the pair operators into angular and amplitude parts
through Eq.\ (\ref{fldecoupling}) where the angular fluctuations that
give rise to the Goldstone modes have been included in the mean-field
action $S_{\rm mf}$ [Appendix A]. 

In a rigorous approach the amplitude fluctuations have to be
integrated out\cite{LV02} in order to determine the contribution of
the action Eq.\ (\ref{SOmega}) to the free energy of the system. In
the light of the {\em local} constraint $|\bOmega_\mu(\br,\tau)|^2 =
1$ the treatment of the resulting terms is not evident. Instead, on
the mean-field level one can estimate the impact of the
amplitude-Goldstone mode coupling by replacing the square of the
amplitude fields by their expectation value, i.e., $\beta
X^2(\bq,i\omega_n) \to \beta \langle X^2(\bq,i\omega_n) \rangle =
\chi^{(P)}_{\rbx{\rm BCS}} (\bq,i\omega_n)$, in Eq.\ (\ref{SOmega}).   

$\chi^{(P)}_{\rbx{\rm BCS}}$ is given in Eq.\ (\ref{chiBCS}) and
diverges for $\omega^2 - v_{\nu}^2 \bq^2 \to \Delta^2$. Consequently
the excitation gap $\Delta$ indeed marks the ultraviolet cutoff
discussed in the literature.\cite{Tsve95,GNT98} Given the mean-field
relation between the excitation gap and the coherence length
$\Delta/\ov_{\rm eff}\approx \xi^{-1}_{\rm BCS}$ the cutoff translates
into a short range or large wavevector cutoff $a |\bq| < \Delta/v_\nu$
in strict analogy to the discussion by Schulz\cite{Schu95} for the
magnetic order in the repulsive Hubbard model.

For sufficiently low temperatures $T\ll \Delta$ the thermal occupation
of states near the cutoff energy can be neglected and the gapless
excitations are well described via the non-linear sigma models given
by Eq.\ (\ref{Snls}) in Sec.\ \ref{sectionfluct}. Since in this case
the amplitude mode has a large gap compared with temperature,
contributions from the amplitude mode and the amplitude-Goldstone mode
coupling term are negligible.   

Closer to the phase transition corrections from Gaussian fluctuations
become relevant\cite{LV02} and the energy and wavevector cutoff will 
play a crucial role and needs to be included in the action Eq.\
(\ref{SOmega}). In continuum representation for the momenta this leads
to Eq.\ (\ref{SOmegaq}) in Sec.\ \ref{sectionSaddle}. Note that for
small energies and wavevectors the $\omega$ and $\bq$ dependence of
$\chi^{(P)}_{\rbx{\rm BCS}} (\bq,i\omega_n) \sim \chi^{(P)}_{\rbx{\rm
    BCS}} (0,0) =  f_0/\Delta$ and $|\bOmega_\mu (\bq,i\omega_n)|^2 \sim
|\bOmega_\mu|^2$ is not expected to alter the physical result and
consequently have been neglected in Eq.\ (\ref{SOmegaq}).


\subsubsection{Limits of the applicability}

It is well known\cite{LV02} that in the critical region around the
phase transition the Gaussian approximation for fluctuation
corrections as derived in Appendix B breaks down. The contributions
are divergent and higher oder terms have to be included in a
non-perturbative fashion. The critical region can be estimated by the
rounding of the measured specific heat near the phase transition (Fig.\
\protect\ref{CP148} in Sec.\ \ref{sectionC}) and yields a temperature 
interval of $T_{\rm c}\pm 0.02$ K. The results herein including Eq.\
(\protect\ref{FLandau}) are only applicable for $T \le T_{\rm c} -
0.02$ K or for reduced temperatures $|t| \ge 0.01$. Note that within
the framwork of the third order phase transition considered in Sec.\
\ref{sectionC} the determination of $T_{\rm c}$ and thus of the
critical region differs from the usual ``mid transition point''
estimate.  

The possible impact\cite{Volo92,Spinspectrumquote,Flavorspectrumquote}
of additional topological terms is addressed in Sec.\
\ref{sectionfluct}.



\begin{thebibliography}{100}
\bibitem{gapquote}
A gap in all channels $\mu$ is also obtained when simply adding the mean-field
  term $H_{\rm mf}$ Eq.\ (\ref{Hmf}) to the high temperature Hamilonians Eq.\
  (\ref{Hc(m)}).
\bibitem{SineGordonSquote}
Here the polar angle has been set $\theta_z = \pi/2$. A description for general
  $\theta_z$ within the framework of the bosonized model is not possible since
  $\theta_z = \pi/2 \to 0$ involves a dual transformation of the spin and
  spin-flavor fields.
\bibitem{Flavorspectrumquote}
Possible low energy contributions with quadratic dispersion\cite{VK83,Volo92}
  are of the order $\Delta^2/v_{\rm F} \sim 10^{-4}$ K and are negligible
  [$\Delta \sim 1$ meV, Sec.\ \ref{sectionOP}].
\end{thebibliography}
\end{document}